\documentclass[12pt,preprint]{aastex}
\shorttitle{Statistical Analysis of  EGRET Blazars}
\shortauthors{Bloom}
\received{2007 March 14}
\begin{document}
\title{The Radio and Gamma Ray Connection of EGRET Blazars: Correlation, Regression, \& Monte Carlo Analysis }
\author{S. D. Bloom \altaffilmark{1}} 
\affil{Department of Physics \& Astronomy, Hampden-Sydney College}
\affil{Box 821, Hampden-Sydney, Virginia 23943}
\altaffiltext{1}{Visiting Scientist, National Radio Astronomy
Observatory, Charlottesville, VA} 
\begin{abstract}
A comprehensive statistical analysis of the broadband
properties of EGRET blazars is presented. This analysis includes sources identified as
blazars in the Sowards-Emmerd publications. Using this sample of 122 sources, we find that there is
a relationship $L_\gamma \propto
{L_r}^{0.77 \pm 0.03} $ as well as a correlation between $\alpha_{og}$ and $\alpha_{ro}$, and a correlation between radio luminosity and $\alpha_{og}$.
Through the use of Monte Carlo simulations, we can replicate the observed luminosity relationship if a synchrotron self-Compton model is assumed. However, this relationship can not be replicated if an external Compton scattering model is assumed. These differences are primarily due to beaming effects. 

In addition it has been determined that the intrinsic radio luminosity of the parent sample falls in the range  $10^{21} < L < 10^{30}\, {\rm Watts\,Hz^{-1}}$ and that the bulk Lorentz factors of the source are in the range $ 1 < \Gamma < 30 $, in a agreement with VLBI observations.

Finally, we discuss implications for GLAST, successfully launched in June 2008.
\end{abstract}

\keywords{gamma rays: observations}

\section{Introduction}
During the lifetime of the Energetic Gamma-Ray Experiment Telescope
  (EGRET) instrument on board {\it Compton
  Gamma-Ray Observatory} (CGRO) 271 sources were detected with 66
  being confidently  identified as blazars in the Third EGRET Catalog \citep{har99}. Many
differing statistical analyses of these gamma ray detected blazars
  have been conducted, among them
\citet{fos98};\citet{muc97}. Some of these analyses concentrated on
the direct
statistical relationship between luminosities in the gamma-ray band and
radio bands \citep{sal96,ste93,fan98}. Of these, most have used single
  dish radio data, but some have used VLBI fluxes \citep{zho97,mat97}, all at various
  radio frequencies $>$ 1 GHz. In addition, \citet{muc97} and \citet{imp96} use
  Monte Carlo simulations to aid in interpreting these
  relationships. Though significant correlations are reported in all
  of these works, \citet{muc97} show that in some cases these will result from
  a combination of variability and selection effects. To
 investigate the variability effects further, \citet{zha01} have compared their
  statistical results using time averaged data for the entire sample to similar results for
  a restricted sample for which data were available during high and
  low states. They report a similar significant correlation in each
  case and show that using time averaged data tends to under estimate
  the underlying linear regression slope (as applied to the logarithmic data).

  Various models have been invoked to explain the origin of the
  gamma-ray emission of blazars and specifically, the radio gamma-ray
  correlation. Among these are the synchrotron self Compton (SSC) model
  \citep{blo96,ghi85} and various models in which the source of seed
  photons for scattering is from a source external to the jet (henceforth called ECS
  models for ``external Compton scattering''). \citet{der93} use the
  accretion disk as the source of soft photons whereas others \citep{ghi96}
  use broad-line region clouds as the source of soft photons.

Since the end of the EGRET mission there have
also been several reanalyses of the significance of identifications of
EGRET sources, particularly those of \citet{mat97,mat01} and
\citet{sow03,sow04}.The \citet{sow03,sow04} survey excludes sources
 with $|b|<10^{\circ}$ (the Mattox papers exclude sources with $|b|<3^{\circ}$),
and thus may exclude additional sources that are thought to be blazars \citep{sgu04}.
Likewise, some work continues on identifying
other individual sources, such as 3EG J0416+3650, possibly identified with
3C 111 \citep{sgu06}. In the eight years since the end of the CGRO
  mission, there has been substantial
modification to the identifications given in the last catalog published by the EGRET
  team \citep{har99}, so we present a comprehensive statistical
analysis including all potential identifications, using homogeneous
criteria for inclusion in the sample ({\S 2}). We have also included various
  statistical techniques of survival analysis for the inclusion of
  upper and lower limits in the data ({\S 3}). We later discuss whether or not the
results of the analyses are dependent on the precise source list.
The approach used in understanding the physical implications of our results is to use a Monte Carlo technique
\citep{lis97} to generate multiple simulated samples under differing assumptions of overall theoretical or
phenomenological model ({\S 4,5}), distribution of physical parameters such as
  bulk Lorentz factor (e.g., Gaussian, power-law, etc.) and selection
  effects. Additional physical implications of observed correlations and implications for the Gamma Ray Large Area Space Telescope (GLAST), successfully launched in June 2008, are discussed in {\S 6,7}.
\section{Data}
To construct a comprehensive list of identified gamma-ray blazars we 
have started with the Third EGRET Catalog \citep{har99}. We note that
we are not re-evaluating the detections themselves. To add or
remove sources from this list of identifications using uniform criteria, we
have further used the guidance of \citet{sow03,sow04}. Though other
standards for inclusion or rejection do exist, we have used these
particular surveys to keep inclusion criterion standard. We have taken
the best blazar identifications with Figure of Merit (FoM) criterion greater
than 0.25 \citep{sow03,sow04}. The FoM used here is a statistic that
incorporates both radio and X-ray spectral information in evaluating
the probability that a particular radio source is a match to the
gamma-ray source. For sources with multiple possible
identifications, we include the identification with the
highest FoM value.  Though the 0.25 cutoff is perhaps arbitrary,
it is clear that well below this limit (at about 0.1), the probability
for chance
associations greatly increases \citep{sow03}.
These authors and their current working group have not yet published an analysis for the sources between -40 and -90
declination. For these sources, we included all of the original
catalog identifications, with the caveat that the more
marginal identifications (i.e. counterparts with 8.4 GHz radio flux
density $< 0.5$
Jy) could possibly not hold up to the analysis using the FoM criterion. The total number of
such sources in this declination range is ten, with two  below 0.5 Jy (the remainder all have
flux density greater than  1 Jy at 8.4 GHz). We also note that we could be excluding candidate identifications heretofore unmentioned
in this declination range. 
A summary of the available data used in our statistical analysis has been compiled in Table 1. Column(1) indicates the
EGRET source name, column (2) indicates the best radio identification,
followed by the redshift in column(3). Columns (4)-(6) give the
monochromatic radio, optical and gamma ray luminosities,
respectively (all have units of $ {\rm joules\, sec^{-1} \,Hz^{-1}}$). If the redshift is unknown, we substitute with z=1. 
There are no significant changes in the results if we use other
assumptions, such as z=0.5. To calculate the radio luminosity we use
the 8.4 GHz flux densities and spectral indices (Column (7))between 1.4 and 8.4 GHz of \citet{sow03,sow04} except for the
lowest declination sources discussed above. For these, we extract the
similar radio data from the NASA Extragalactic Database (NED) to calculate the spectral index between
1.4 and 8.4 GHz. Henceforth we use ${F_{\nu}} \propto {\nu^{-\alpha}}$
for defining spectral indices. In calculating the optical luminosities we use
recent V magnitudes extracted from NED (when available), and we derive an
optical flux density and monochromatic luminosity from this magnitude using the conversions of \citet{bes79}.
\citet{sow03,sow04} also provide archival R and B magnitudes from the
USNO catalog \citep{mon03}, which we have used to interpolate a V value if we did
not otherwise have a reference for a V magnitude. In a few cases we
only had O magnitudes from the POSS-I survey or 2MASS infrared
magnitudes, and thus need to extrapolate to the V band.
For sources that
have no optical or near infrared magnitudes in the literature, and
that were not detected in the POSS-I, SERC-J, or POSS-II we
assume magnitude upper limits equal to the appropriate plate limits of
the survey (eg., O=21.5,E=20 for POSS-I \citet{mcm02}) 
and
derive upper limits to the optical luminosities accordingly.
The gamma-ray luminosities are determined from
the flux and spectral index (Column (8)) given in \citet{har99}. We
use the formula of \citet{tho96} to
derive an exact 400 MeV flux. If the spectral index is not
known, we assume it is 1.0 (note that we are referring to the energy
spectral index and not the photon index). In all cases we use the co-added flux
over all viewing periods. In the absence of simultaneous data at all
wavelengths, this is a better representation of broad band
properties than any single value. Each of the luminosities is
calculated using equation (2) of \citet{lis97} and assumes a flat $\Lambda$ cosmology with ${\Omega_m}=0.3$
and ${\Omega_{\Lambda}}=0.7$
\citep{per97}. Though the luminosity distance can be calculated via numerical integration, it is preferred to find 
an analytic expression, especially for use in the Monte Carlo
simulations where such a calculation will have to be performed many
millions of times. \citet{pen99} offers such an option; however
this formula can differ from the numerically determined value by over
5\%. Therefore, a polynomial  fit to the numerical solutions is
determined for the redshift range z=0-5 using the Wolfram Research program
{\it Mathematica} to attain a more accurate approximation. When luminosity distances determined via this polynomial
fit were compared with the results of a direct numerical integration, the
agreement was well within 1 \% for the range $ 0.1 < z <4.0 $; however, at redshifts outside of this range, the disagreement is as large as 7\%.
 Column(9) shows the broadband
spectral index, $\alpha_{ro}$ 
gives the index 
$\alpha_{og}$. 
Column (11) has the gamma-ray variability parameter, $\delta_{var}$,
as determined by \citet{nol03}. Column (12) gives the references for
the data, where the first reference is for redshift, the second for
the radio data, the third for the optical magnitudes, and the fourth for
the gamma ray data.
\section{Statistical Analysis}
We have evaluated the strength and significance of correlation between
monochromatic luminosities using non-parametric methods, such as
Kendall's $\tau$ \citep{ken90}. These results are given in
Table 2. Column(1) gives the independent variable, Column(2) the
dependent variable, Column(3) the number in the sample,
Column(4) Kendall's $\tau$ statistic, Column(5) the probability of
the null result for this statistic, Column(6) the Spearman's $\rho$ statistic,
Column(7) the probability of the null result for this statistic, and in 
Column(8) the
regression technique used. Here BJ refers to the Buckley-James method,
EM refers to the EM method and SB refers to Schmitt's binned
regression, as explained in the references that follow.
Column(9) gives the linear regression slope, and
Column(10) the linear regression Y-intercept. In some cases, we have upper and lower limits to measured
values, and to take this into account, we use the proper correlation
and regression techniques as discussed in \citet{fei85,iso86} and utilized in the ASURV
software package. In reporting correlation coefficients, probabilities and regression fits we adopt the format of \citet{muc97}. That is, we report three significant digits down to 0.100. Between 0.010 and 0.100 we report two significant digits, and for all numbers with values below 0.01, we only report with 1 significant digit.
 
Most notably, we determine that there is a strong
correlation between gamma ray and radio luminosity (Figure 1 and Table
2). In order to
determine whether the correlation is affected by including the sources with
FoM $<1$ (the lower confidence identifications), we recalculate the correlation and regression coefficients
while retaining only the 76 sources with FoM $>1$. The strength of the
correlation is similar, with a very slight steepening of the regression
slope (0.78 instead of 0.77, which is within the range of the
uncertainty of the correlation slopes). If the sources with declination below -40 $^\circ$ are also excluded, the the correlation slope flattens slightly to 0.73. In all though, correlation is not significantly altered by such changes to the source list.

For some correlations, we must account for the effect of a third
variable, using partial correlation analysis \citep{pad92}. Results of our partial correlation analysis 
are given in Table 3. Column(1) gives the independent variable, Column(2) the dependent variable,
Column(3) the third variable, Column(4) the number in sample, Column(5) Kendall's $\tau$, and
Column(6), the probability of the null result. Examples of this are the need to take into account the correlation of
all luminosities on redshift, and the dependence of several broad band
spectral indices on optical luminosity. In particular we note that the
strong correlation between gamma ray and radio luminosity is
significantly weaker, though still significant, once we account for
the partial correlation with redshift. Similarly, the correlation
between the $\alpha_{ro}$ and $\alpha_{o \gamma}$ still persists, even
after the effect of the common dependence on optical luminosity is
negated. We use the Kendall's $\tau$ coefficient for this purpose, since
to date it is the only coefficient for which there are methods to take
into account censored data \citep{akr96}. A detailed discussion of the
physical models for the luminosity correlation is addressed in {\S 4,
 \S 5}. Further discussion of the remaining correlations is deferred
to {\S 6}. 
\section{Physical Models}
Our approach in this section is to see if particular models can be used to
explain our statistical results above, particularly the radio and gamma-ray luminosity relationship.
A radio and gamma-ray luminosity correlation would naturally be
expected given any of the non-thermal inverse Compton models linking
low energy emission with gamma rays, such as in \citet{blo96}. As
noted above, this luminosity correlation
holds even after we take the effect of redshift into account. A linear or nearly linear correlation is possible within the synchrotron self-Compton
(SSC) model if the range of possible physical parameters, particularly the
optically thin spectral index, Thomson optical depth  and electron
energy cutoffs do not vary
too greatly in the sample. This can be illustrated by following
\citet{mar87} for a spherical homogeneous source:

$$ {L_{\nu c} \over L_{\nu s}}=
{{c_3(\alpha)}{N_0}R ln{({\nu_m \over \nu_2})}{{( {\nu_c \over
	\nu_s})}^\alpha}}\eqno {(1)} $$

Where $L _{\nu c}$ is the frequency dependent Compton luminosity, $L_{\nu s}$ is the
frequency dependent synchrotron luminosity. $c_3$ is a function of
spectral index, tabulated in \citet{mar87}, $N_O$ is the normalization
factor of the energy dependent electron density distribution, and $R$
is the radius of a spherical source. $\nu_m$ refers to the spectral turnover frequency
in a plot of flux density versus frequency, and $\nu_2$ is the upper frequency cutoff
in a similar plot. A correction to this formula needs to be applied if the observed gamma-rays are 
at a frequency beyond the high end spectral cutoff for Compton scattering, or if the synchrotron photons are observed at
a frequency at which the source is believed to be optically thick or partially opaque.
The logarithmic term only varies a small amount for large changes in the parameters,
so this term can usually be neglected relative to the others.
The function of $\alpha$ can vary over several orders of magnitude,
but can be roughly constant if the optically thin spectral index of
the sample is narrowly distributed. The factor of ${N_O \,R}$,
proportional to the Thomson optical depth, would also have to be
tightly constrained. Using a somewhat different
formulation, \citet{ghi98} calculate a  particle injection
compactness parameter, which is also proportional to the Thompson optical
depth. They show that this parameter, though distributed over 4 orders
of magnitude for a
sample of 51 EGRET blazars, has a
clear peak value (over half of the blazars with a value within 1 order
of magnitude of the peak). Thus, in adopting  a linear model for use
in the following Monte Carlo analysis, this parameter can
be drawn randomly from a sharply peaked distribution.   
It is not as evident that these arguments can be extended to
a relationship such as that observed, ${L_{\gamma}} \propto {{L_r}^{0.77}} $. We explore this possibility below.

Alternatively, the gamma-ray luminosity can be produced by an
external Compton scattering process
(ECS)\citep{der93,lis99a}. In this particular model the source of the soft
photons is the accretion disk, and these photons scatter off of a plasma of
relativistic electrons in a blob. Adapting the formulas of \citet{der92} to
our notation:

$${L_{\nu c} \over L_{\nu,acc}}=
{ {R^3\over 6r^2} {\sigma_T}{(m_ec^2)^{-2\alpha}}{N_0}{\delta^{2\alpha+4}}{(\nu/\nu_o)^{-\alpha}}}$$

Here, the luminosity ratio is between the observed Compton luminosity
in the gamma-ray range and the observed accretion luminosity. $R$
refers to the radius of the blob, assuming it is spherical, and $r$
refers to the distance between the blob and the accretion
disk. $\sigma_T$ refers to the Thomson scattering cross-section,
${m_e}c^2$ refers to the electron rest mass, $N_0$ has the same meaning as above. $\delta$  is the Doppler factor
to correct for bulk relativistic motion of the emitting plasma and is
defined by:

$$\delta \equiv {1 \over {\Gamma(1-\beta cos\theta)}} \eqno{(3)}                                          $$
Here $\beta$ has the usually meaning of ratio of bulk speed to that of light in a vacuum, and $\theta$ is the angle between direction of the bulk flow and the line of sight of the observer.
The power to which the Doppler factor is raised is dependent on the structure of the emitting region (discrete blobs vs. continuous jet). The formula above is correct for discretely emitting blobs. For a continuous jet, the power is reduced by one \citep{der95}.
It has been suggested that the 
synchrotron luminosity can be proportional to the ultraviolet luminosity
of the seed photons \citep{ghi96}, thus providing a direct relationship
(though not necessarily linear) between the gamma-ray and radio luminosity within the framework of the
ECS model. Furthermore, the underlying radio and gamma-ray correlation may
also be explained by other means, since the gamma-ray luminosity in
all relativistic Compton scattering models is proportional to the
Thomson optical depth and the intrinsic
radio luminosity can be proportional to the Compton optical depth via
the parameter $N_0$ and $R$. Other source parameters,
such as the magnetic field, would have to fall into a narrow range for
the sample in order for this correlation to hold.
\section{Monte Carlo Simulations}
The goal of our Monte Carlo simulations is to generate the
luminosities and flux densities of a hypothetical sample that,in principle,
mimics the sample of gamma-ray detected blazars discussed above. To do
that we need to define a radio luminosity function from which we will
randomly draw to create the sample. We will also need a distribution of Lorentz factors
which will be used in determining Doppler beaming factors needed to
transform between  observer frame luminosities and intrinsic luminosities.
In addition, we will need a model that determines what the gamma-ray
luminosity will be for a given radio luminosity. We will limit
ourselves to the models discussed above. Lastly, we will need to adopt
a cosmological model in order to determine redshifts corresponding to randomly
chosen enclosed volumes of space. We can then determine both radio and gamma-ray flux densities using these redshifts and the adopted cosmological model. 
Following \citet{lis97} we have used an
assumed radio luminosity function of the form:

$$ \rho(L) \propto L^{-g}\eqno{(4)}$$ 

valid between $L_1$ and $L_2$. We have adopted a luminosity evolution function:

$$L(z)=L(z=0)exp[T(z)/\tau] \eqno{(5)} $$

Where $T(z)$ is the look-back time and $\tau$ is a constant in the approximate range of 0.1-0.5 that determines the degree of evolution.
 
We have also assumed
a distribution of Lorentz factors for which:

$$N(\Gamma) \propto \Gamma^{-s} \eqno{(6)}$$

valid between $\Gamma_1$ and $\Gamma_2$ and a linear model connecting
gamma-ray and radio luminosities, i.e., the inverse Compton models discussed above,
to generate a simulated sample with Monte Carlo techniques and then use the corresponding distributions 
(luminosity, redshift,flux density, etc.)  and correlation diagrams to assess
the validity of the model. We express these
models using the following equations :

$$ L_{\gamma,o,ssc}=K_{ssc}L_{r,i}\delta^{3+\alpha} \eqno{(7)}$$
$$ L_{\gamma,o,ecs}=K_{ecs}f(\beta,\theta)L_{r,i}{\delta^{4+2\alpha}}\eqno{(8)}$$

Here, $L_{\gamma,o,ssc}$ and $L_{\gamma,o,ecs}$ refer to the predicted
gamma-ray luminosities in the observer's frame, using the SSC and ECS
models respectively. The K constants depend on the parameters
discussed above, $L_{r,i}$ refers to the intrinsic radio luminosity,
and the remaining factors are the  appropriate functions of Doppler
factors for SSC or ECS assuming that the radiation is from discrete blobs \citep{lis99a}. As mentioned above, the powers in each case are reduced by one for the case of a continuous jet.

\citet{lis97} rigorously determines radio luminosity function parameters based on simulations of the 
Caltech-Jodrell Bank sample (CJ-F, as labeled by \citet{lis97}), taking into account distributions of
bulk Lorentz factors, redshifts, and luminosities. \citet{car07} have
recently arrived at comparable results for the similar  MOJAVE sample
(though the luminosity and redshift ranges differ somewhat). We
begin by adopting the parameters that \citet{lis97} derived for the
CJ-F sample. We expect the CJ-F parent population to be very similar to that of our sample, because it is selected for flat spectrum, compact structure and 5 GHz total flux density $>$ 0.35 Jy. Additionally we take into account beaming effects appropriate for the SSC or ECS
model. Then, assuming both SSC and ECS in
turn, the gamma-ray luminosity and flux density is determined for each source. 
In generating the gamma-ray luminosities from the intrinsic radio luminosities
we assume a power-law distribution of the K values discussed
above in Equations 7 and 8. This is an arbitrary choice of distribution, but the precise values of power-law
slope and cutoffs do not significantly effect the results.
We then generate 122 sources with gamma-ray flux that could have been
detected by EGRET. We adjust the gamma-ray flux density limit
to be slightly lower than that of EGRET (about 85 \%) to allow for a
proportion of upper limits that is similar to that we see in our observed sample.
In looking for agreement between the simulated
population and the observed one, we use the two distribution
Kolmogorov-Smirnov test on the distributions of luminosity, redshift
and flux density. The $D_{KS}$ statistic is then used to measure the
discrepancy between the two distributions. $D_{KS}$ is defined as \citep{pre86}:

$$ D_{KS} \equiv  { {\rm max} \vert S_{N1}(x)-S_{N2}(x)\vert } \eqno{(9)} $$

That is, $D_{KS}$ is the maximum difference (absolute value) between the two distribution functions, $S_{N1}(x)$ and $S_{N2}(x)$. The distribution function is defined to be the probability that the particular value of the parameter in question is less than x. A value of $D_{KS}$ near zero indicates
agreement between the distribution functions, whereas a value near 1 indicates a large discrepancy. The
associated probability is also determined following \citet{pre86}. In this case, a probability approaching 1 indicates a high
probability for agreement between the distributions. For distributions
that include upper limits we use the corresponding tests using the
ASURV program (this is only an issue for the gamma ray flux densities
and luminosities).Among these tests are the Gehan generalized Wilcoxon
test and the Peto \& Peto Generalized Wilcoxon test. The results of
these tests are discussed below.
We have applied these methods using
the parameters of \cite{lis97}, summarized in Table 4.
In Table 4, Column (2) gives $L_1$, Column(3) gives  $L_2$  and they refer to the lower and upper
luminosity limits of the intrinsic luminosity function. Column (4) gives
g and refers to
the luminosity function power law slope. Column(5) gives $\Gamma_1$,
Column(6) gives $\Gamma_2$ and 
Column(7)gives s. These labels refer to the lower bulk Lorentz factor
distribution limit, upper bulk Lorentz factor distribution limit and the slope of the power law,
respectively. Column (8) gives $\tau$ and in this case refers to the evolution parameter, assuming an
exponential evolution model (see Equation 5). Column(9) gives $z_{min}$ and is the
lower limit to the redshift distribution and Column(10) gives  $z_{max}$ and is the upper limit. Column (11) gives $h$ as discussed in {\S 2}, Column (12) gives the minimum value of $K$, Column (13) the maximum value of $K$ and Column (14) gives the slope of the power-law distribution of $K$, where $K$ refers to either $K_{ssc}$ or $K_{ecs}$ (Equations 7 and 8), whichever is appropriate for the simulation indicated.  Applying the values assumed by \citet{lis97} does not
immediately lead to good agreement between the observed and modeled
distributions (see Table 5). For models marked "Lister SSC" we have adopted
an SSC model to generate simulated gamma ray luminosities and for "Lister ECS" we have adopted an ECS model to generate simulated gamma ray luminosities. 
Next we have explored parameter space to determine where
better models might lie. Since the number of parameters and the
possible range of parameters is large, there are likely to be
multiple models that would have similar agreement as seen by a K-S test.
The results that give the lowest value of $D_{KS}$ (Equation 9) with high significance  are summarized
in Table 5 for the SSC and ECS models. Column(2) gives the $D_{KS}$ value
for the radio luminosity function and Column(3) gives the associated
probability that the distributions are drawn from the same parent
population. Column(4) and Column(5) give the same results for the
gamma ray luminosity distributions, whereas Columns (6) and (7) do so
for redshift, and Columns(8)-(11) do so for the radio and gamma-ray
flux density distributions, respectively. These results can also be
visually inspected in Figures 2-11.  There is only weak agreement
between the observed and predicted distributions for either SSC or ECS.
For the special case of the gamma-ray flux and luminosity
distributions (which have limits as well as measured values), we have
used the tests discussed above, and generally the associated
probabilities (for agreement of the two distributions)  for these
tests fall in the range of 0.0002-0.0015 for SSC and 0.0001-0.0010 for ECS, thus also only very weakly confirming
the SSC or ECS hypothesis. The parent population in these simulations was approximately 30
million for SSC and 300 million for ECS. Both models require a range of
Lorentz factors of about 1-30. This is the approximate range inferred by VLBI observations  \citep{coh07}. However, some authors have made the argument that these bulk Lorentz factors may not pertain to the same portion of the jet from which the gamma ray
emission originates \citep{geo03}.

A closer look at Figures 2-11 show some features that are not as evident from just comparing results of Kolgomorov-Smirnov tests. A comparison of the radio luminosity distributions (Figures 2a,2b,\& 7) shows that assuming an SSC or ECS model lead to similar results. Both the SSC and ECS distributions have peak values near $10^{27} {\rm Watts \, Hz^{-1}}$, with approximately 24 \%  of the sample having values near the peak for SSC and 21 \% for ECS. The observed distribution of radio luminosity peaks near $10^{28} {\rm Watts \,Hz^{-1}}$ with approximately 33 \% of the sample in the bin nearest to the peak. However, in conjunction with the redshift and flux density distributions we do see some important differences between the predictions of the SSC and ECS model, and can pin point particular observations that will be necessary in the GLAST era for differentiating these models using the methods we have discussed above.     
For instance, when looking at the simulated distributions of radio flux density (Fig. 4a, Fig. 9 ), the simulated distribution assuming the SSC model is clearly peaked close to about 0.3 Jy, whereas the same distribution for ECS is peaked close to 1.5 Jy. The observed distribution is peaked closer to 0.3 Jy and is more similar in overall shape to the predicted SSC distribution (Fig. 4b). Ongoing radio monitoring of GLAST detected sources (or at least updated to these older survey flux densities) will be critical for determining the true observed distribution to be used for comparison to models. In addition, we can see that the redshift distribution (Fig. 6a) with SSC assumed is nearly identical to the corresponding distribution with ECS assumed (Fig. 11). Though the observed redshift distribution (Fig. 6b) is also peaked near z=1, this is at least in part due to the assumed redshifts of z=1 for 13\% of the sample.  An additional challenge in th GLAST era will be obtaining re!
 dshifts for entire large samples (possibly thousands of new quasars that have not yet been optically identified or observed spectroscopically). It is a bit harder to discern significant differences between the ECS and SSC predictions for either gamma-ray luminosity or gamma-ray flux density. However, a significant increase in sensitivity for GLAST should clarify the nature of the low-end of each of these diagrams. For example, we now can only see
a very large drop-off in all of the flux densities near $10^{-11}$ Jy. This is caused by the flux detection limit of EGRET. In addition, the dramatic increase in number of gamma-ray detected sources to thousands \citep{pad07} will clarify the shape of of these distributions and increase the conclusiveness of comparisons to theoretical predictions.

We have also used the Monte Carlo results to generate simulated
luminosity-luminosity correlation plots between $log L_{\gamma}$ and $log L_r$ by assuming, in turn, the SSC and
ECS models. The SSC simulations generate the relationship $L_{\gamma} \propto{L_r}^{0.74 \pm 0.02}$, in agreement with observations. However, assuming the ECS model generates results with $L_{\gamma} \propto
{L_r}^{0.95 \pm 0.02}$ (the results are plotted in Figures 12a and 12b). This clearly shows that though the assumed underlying physical models in both the SSC and ECS case are linear  between $L_{\gamma}$ and $L_r$, selection effects can lead to a simulated sample that shows a non-linear relationship between the luminosities. This effect is more pronounced for the SSC case. 
This is potentially  an effect caused by Doppler beaming factors raised to different exponents (dependent on spectral shape). For instance, in the SSC case, if the exponents for the beaming factors are identical for both the radio and gamma ray part of the spectrum, then we would expect no change in going from the intrinsic luminosity correlations and the observed ones. However, in our model we do account for the possibility that the gamma-ray spectrum is considerably steeper than the radio spectrum. This leads to somewhat different exponents for the Doppler beaming factors that are going into an overall beaming correction that is 
applied to the intrinsic radio and gamma ray luminosities. There are two additional effects which we have not modeled, but may contribute to the observed luminosity correlation. In particular, the radio source and gamma-ray source may not be spatially coincident, and may thus have different bulk Lorentz factors, as discussed above. Also, the bulk Lorentz factors may be dependent on radio luminosity \citep{lis97}.  

Though similar effects are also at work with ECS process, it is less pronounced in our case due to a selection effect. That is, because the range of $K_{ecs}$ is very narrow, and also has relatively low values, only the sources with the highest values of Lorentz factors are detected in gamma rays. Thus, relative to the SSC case, the Doppler factors are closer to being constant for the sample. Overall, this leads to a relationship between the gamma-ray and radio luminosities that is closer to being linear. 

\section{Additional Correlations}
In addition to the luminosity correlation that we have discussed at
length we have also observed other significant correlations for the sample.
The origin of the broad band spectral index correlation, $\alpha_{og}$
vs. $\alpha_{ro}$ (Table 2), becomes clear if we consider the conclusions of 
\citet{fos98}. They show that gamma-ray dominance (the ratio of gamma-ray luminosity to luminosity at the 
model dependent synchrotron peak of the SED) is positively correlated with the maximum
frequency of the lower energy peak of the SED. In a simple
homogeneous source, this lower energy peak corresponds to the
upper synchrotron cutoff frequency discussed above. \citet{ghi98} show that this correlation is a proxy for
the underlying physical relationship between the maximum relativistic electron energy (on which the low frequency peak strongly
depends) and the ratio of the Compton and synchrotron luminosities. For a model in which both an internal (SSC) 
and external (ECS) radiation field are present, an increase in the energy density of the external field leads to 
an increase in Compton cooling and thus a decrease in the maximum electron energy and an increase in the Compton gamma-ray
luminosity \citep{ghi98}. 
\citet{fos98} further show that observed optical flux density can be substituted for the (model dependent) synchrotron peak flux
density without loss of the Compton dominance/spectral peak frequency correlation. We utilize this useful substitution 
below.
\citet{fos98} also show that both $\alpha_{rx}$ and
$\alpha_{ro}$ are separately but similarly correlated with this same spectral maximum frequency, so that the
latter may be substituted for the former in the absence of X-ray data.
Though we have limited spectral data for many sources, and thus 
can't confirm this related correlation for our own sample, the same
reasoning should apply to our sample, and thus we make this substitution 
as well. Thus the correlation of $\alpha_{og}$ and $\alpha_{ro}$ is a
possible indicator of the relationship between dominance of gamma-rays
produced by the ECS process and maximum electron energy.
However, it is also possible that the broad band spectral index correlation
is not independent of the radio/gamma-ray luminosity correlation
discussed above. This is especially true if we take into account the
common dependence of both spectral indices on optical luminosity
(Table 3). Recent surveys have also suggested that the aforementioned correlations discussed in \citet{fos98} and \citet{ghi98} may be due in part to selection effects \citep{ant05,nie06}. For instance, the low luminosity objects of \citet{fos98} were mainly X-ray selected, favoring selection of objects with SEDs peaked at higher frequencies, whereas the high luminosity objects were mainly radio selected, thus favoring the selection of objects with SEDs peaked at lower frequencies \citep{ant05}.    

The correlation between radio luminosity and gamma-ray dominance
(which in our analysis would be characterized by $\alpha_{og}$) ,
 originally seen by \citet{fos98} is also seen here, after more than doubling the sample size.  Here, because of 
the way the broad band indices are defined, a lower value of $\alpha_{og}$ corresponds to greater gamma-ray dominance.
This correlation is further evidence for a blazar sequence defined by relative importance of the ECS model
at higher luminosities.

We have also examined a possible correlation between all parameters
individually and the gamma-ray variability \citep{nol03} and find 
only a weak anti-correlation between gamma-ray spectral index, $\alpha_{\gamma}$, and variability (as parameterized by \citet{nol03}; see Table 1). This may be related to a spectral hysteresis effect observed by \citet{nan07} for a sample of 26 particularly bright and well observed blazars. These authors find that during gamma ray flares the spectral index first tends to flatten with increasing flux, and then returns to a steeper index as the flare ends. Thus, it is plausible that if a blazar is more highly variable there will be an increasing chance of observing it while its spectral index is flatter than for other blazars that are not gamma ray variables.
\section{Conclusions and Implications for GLAST}
After taking into account statistical tests and Monte Carlo analysis,
we find the following:
\begin{enumerate}
\item For this sample of 122 gamma-ray blazars there is a strong
  correlation between radio and gamma ray
  luminosity which persists even after the effects of redshift and
  limits are taken into account. The correlation is of the form 
  $L_{\gamma} \propto {L_r}^{0.77}$. This correlation remains with
  similar regression coefficients even when only the strongest
  76 candidates are included in the sample.
\item There is a correlation between $\alpha_{og}$ and $\alpha_{ro}$
  as well as a correlation between $L_r$ and $\alpha_{og}$. Each
  correlation is consistent
  with the increasing dominance of the SED by gamma ray luminosity as 
the maximum relativistic electron energy decreases and the increasing
  importance of the ECS process at high radio luminosities and low
maximum electron energy.
\item A detailed simulation of source statistics using Monte Carlo
techniques shows that the relationship $L_{\gamma} \propto {L_r}^{0.77}$ can only be reproduced assuming the SSC model. Though the  assumed intrinsic physical model in the source frame is of the form $L_{\gamma} \propto L_r$, selection effects lead to a simulated sample with the relationship described above. This effect is much less significant when ECS is assumed as the underlying model.  However, upon also comparing the observed and simulated distributions of the luminosities, flux densities and redshifts, both the SSC and ECS models are only weakly consistent with the data, assuming linear dependence of intrinsic
  gamma ray luminosity
  on intrinsic radio luminosity. Taken together with the previous
  results, this would suggest that if either SSC or ECS is indeed
  responsible for the gamma ray emission, a more complex model is
  likely needed. In addition, effects of evolution and assumed
  cosmologies can be explored in more detail.
\end{enumerate}
In the GLAST era, our findings can be clarified in the following ways. Acquiring more gamma-ray data, including, particularly, detections of new dim sources near the GLAST detection limit will clarify whether the correlations we find are caused by truncation effects due to the flux limits of our instruments, or whether this is due, at least in part, to physical causes, at least down to the new limit established by GLAST. 

In order to predict what we might see with GLAST, we can extend our observed radio/gamma ray luminosity correlation down to lower luminosities than what are covered in Figure 1. A radio luminosity of $1.25 \times 10^{24} {\rm Watts\,Hz^{-1}}$ would lead to a gamma ray luminosity of approximately $4.98 \times 10^{13} {\rm Watts \, Hz^{-1}}$. We can then covert these luminosities to flux densities  for a range of potential redshifts. For z=0.1-1.1 the radio flux density would be in the range of 0.8-60 mJy and the gamma-ray flux density would be in the range,
$3.3 \times 10^{-14}$-$2.4 \times 10^{-12}$  Jy. Greater values of redshift would lead to even lower flux density values. Most of the range of radio flux densities ($>$ few mJy) would be detectable by the Green Bank Telescope (GBT), with the lower limit detectable with the Very Large Array (VLA)\citep{con08}. Assuming that the sensitivity of GLAST is approximately $2.8 \times 10^{-13}$ Jy at 100 MeV, these sources would only be  be detected if  $z< 0.2$. A radio luminosity of $1.58 \times 10^{23} {\rm Watts \, Hz^{-1}}$ corresponds to a gamma-ray luminosity of $1 x10^{13} {\rm Watts \, Hz^{-1}}$ . The radio flux density would fall in the range of 0.1-8 mJy for z=0.1-1.1. These sources could potentially be  radio detectable with  the VLA under optimal conditions. At the higher flux density limit it would also be possible to detect with GBT, especially at higher frequencies (i.e., 40-50 GHz)\citep{min08}. However, if the regression slope were actually closer to 1, implying a d!
 irect proportionality between radio and gamma ray luminosities, then radio sources in the same luminosity, redshift and flux ranges discussed above would not be detected in gamma rays at all. In short, for the observed luminosity relationship for EGRET blazars to be appreciably extended down to lower luminosities with GLAST detections, then a large percentage of GLAST sources near the detection limit would have radio flux densities about 10 mJy or greater (and $z< 1$).  Several authors have stated that the predicted number of blazars to be detected by GLAST rougly matches with the sky density of flat spectrum radio sources down to 50 mJy at 5 GHz \citep{pad07} and 65 mJy at 8.4 GHz \citep{hea07}. However, our results show that a large number of GLAST sources would be detected at even lower radio flux densities if the luminosity relationship we observe for EGRET sources also holds up for GLAST sources. It is very likely that many new radio observations would be required in a!
 ny of the cases mentioned above, but especially for cases in w!
 hich the
 putative radio source may not even be in any previous catalog.  

>From a theoretical perspective, confirmation of the previously determined correlation between gamma-ray and radio luminosities will lead to confirmation of SSC, though the precise agreement would have to be re-analyzed with these new data. 

\acknowledgements
The author thanks M. Lister for many useful discussions and the
anonymous referee for abundant constructive comments.
S. Bloom would like to acknowledge the generosity of the National Radio
Astronomy Observatory (NRAO) during his stay as a Visiting Scientist at NRAO
Headquarters in Charlottesville, VA. S. Bloom would also like to acknowledge several Summer Faculty Fellowship grants from Hampden-Sydney College.

This work has also made extensive use of the NASA Extragalactic Database
(NED) and NASA's Astrophysics Data System Bibliographic Services (ADS).

{}

\clearpage

\begin{figure}
\figurenum{1}
\plotone{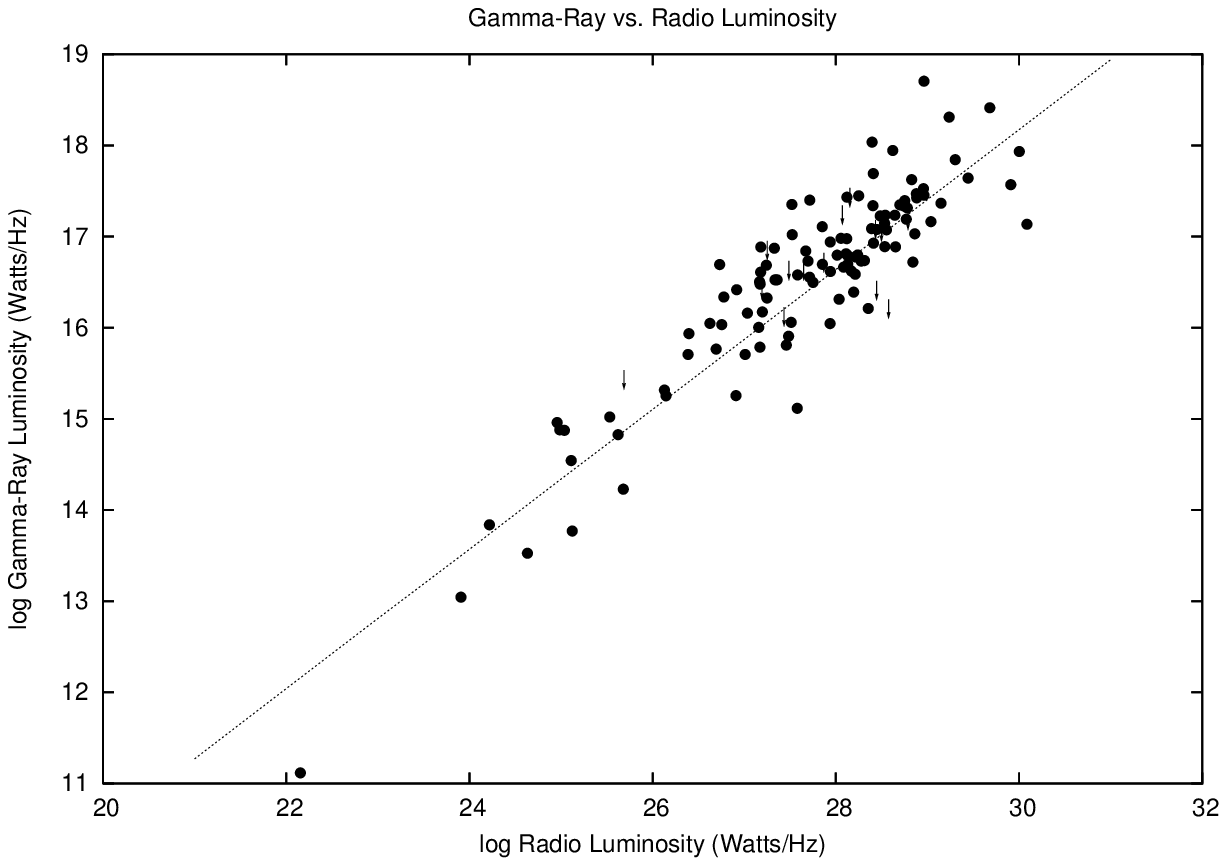}
\caption{The observed radio/gamma ray luminosity correlation. Arrows
  indicate upper limits and the dashed line represents the
  regression fit.}
\end{figure}
\begin{figure}
\figurenum{2}
\plottwo{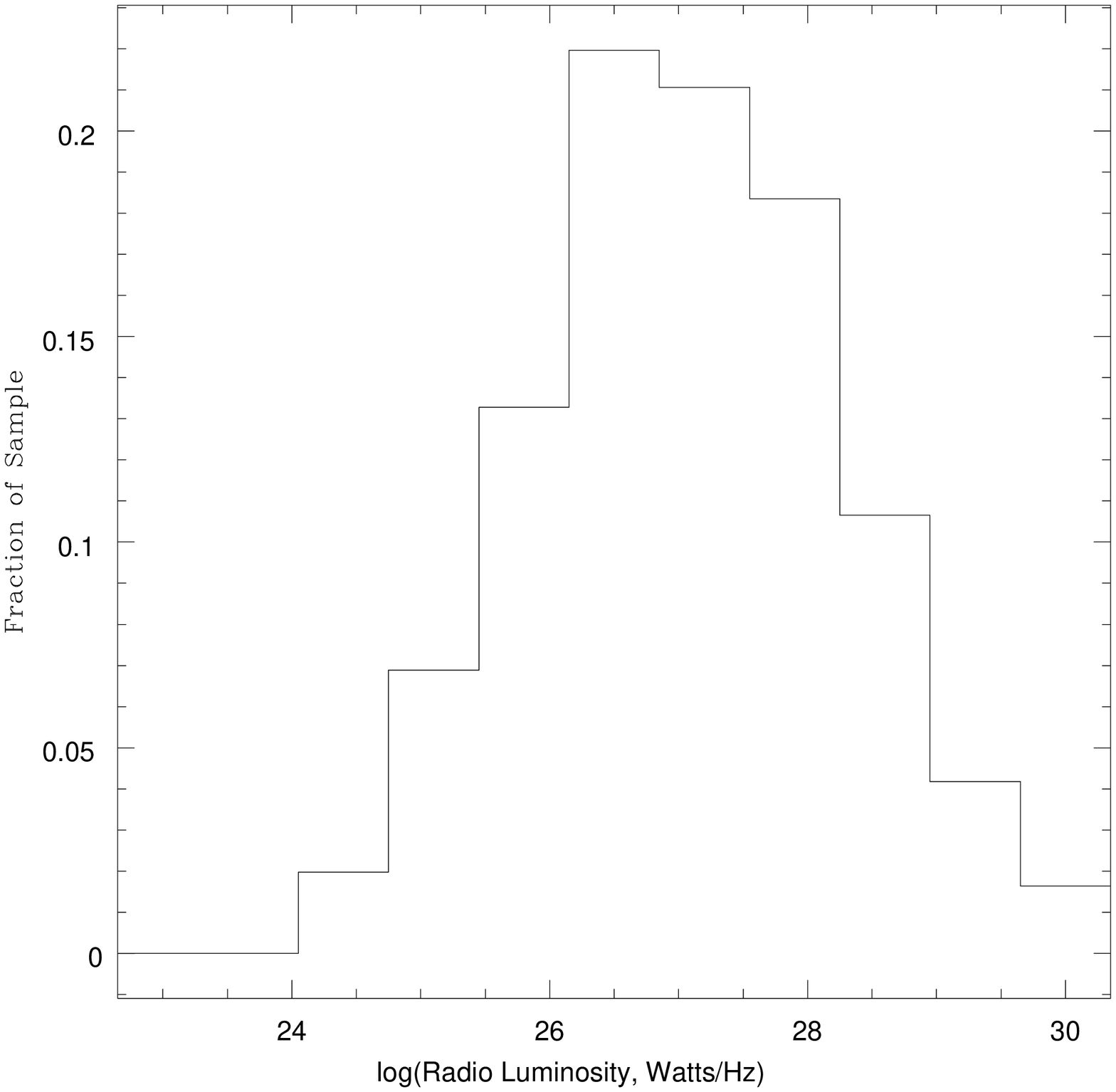}{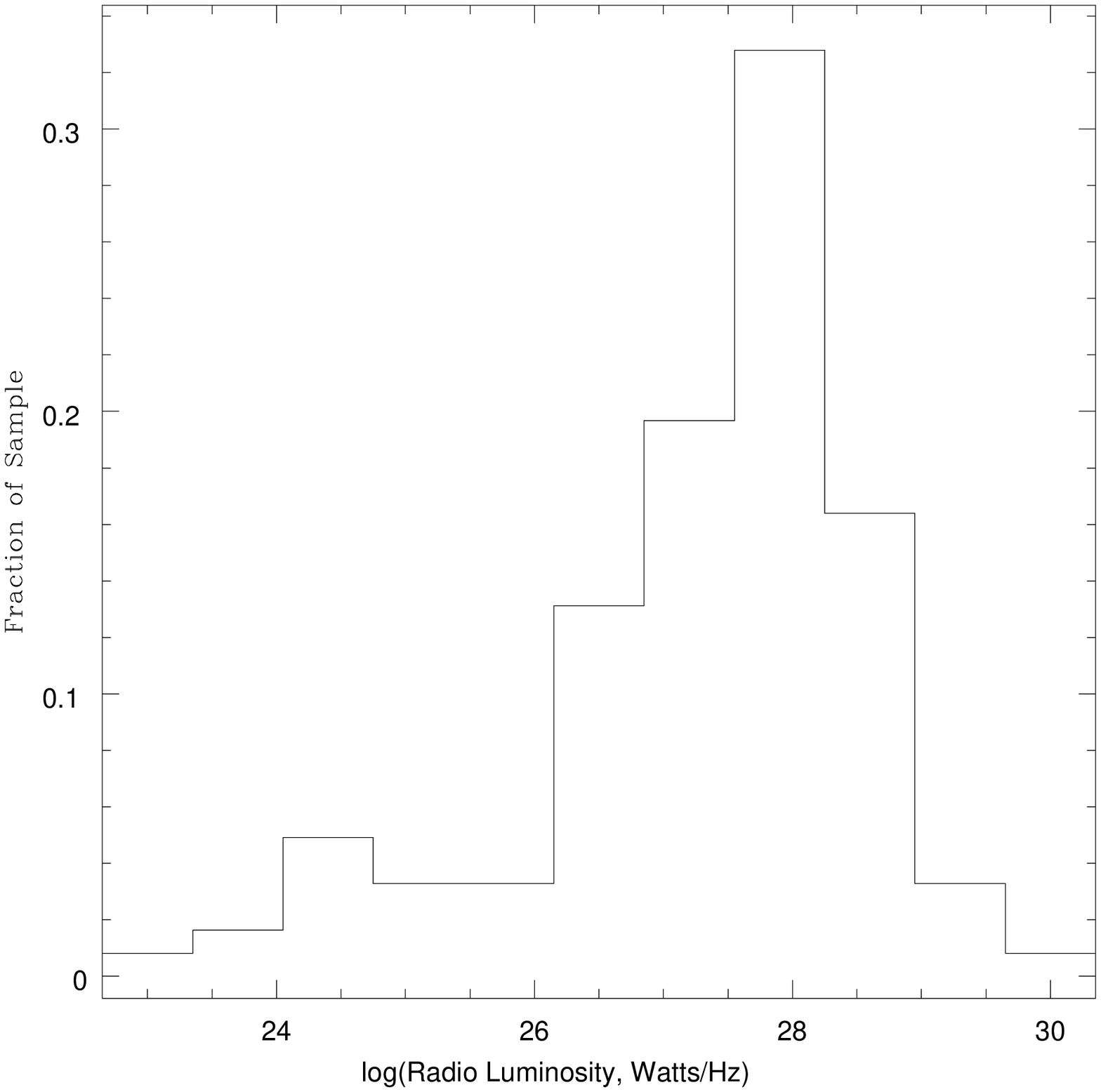}
\caption{Simulated (left) and Observed(right)  Radio Luminosity
  Distributions (SSC model assumed for simulations)}
\end{figure}
\begin{figure}
\figurenum{3}
\plottwo{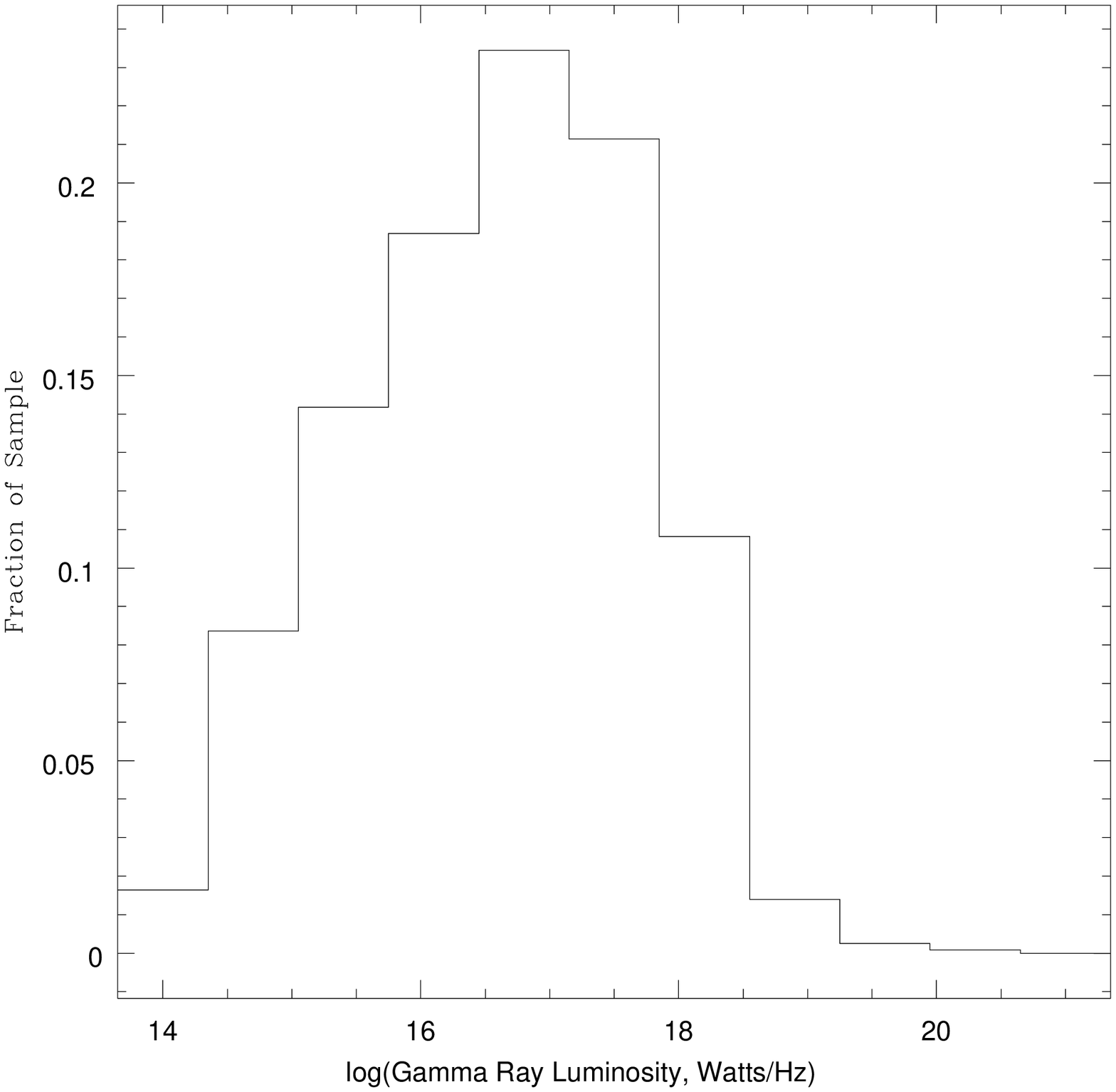}{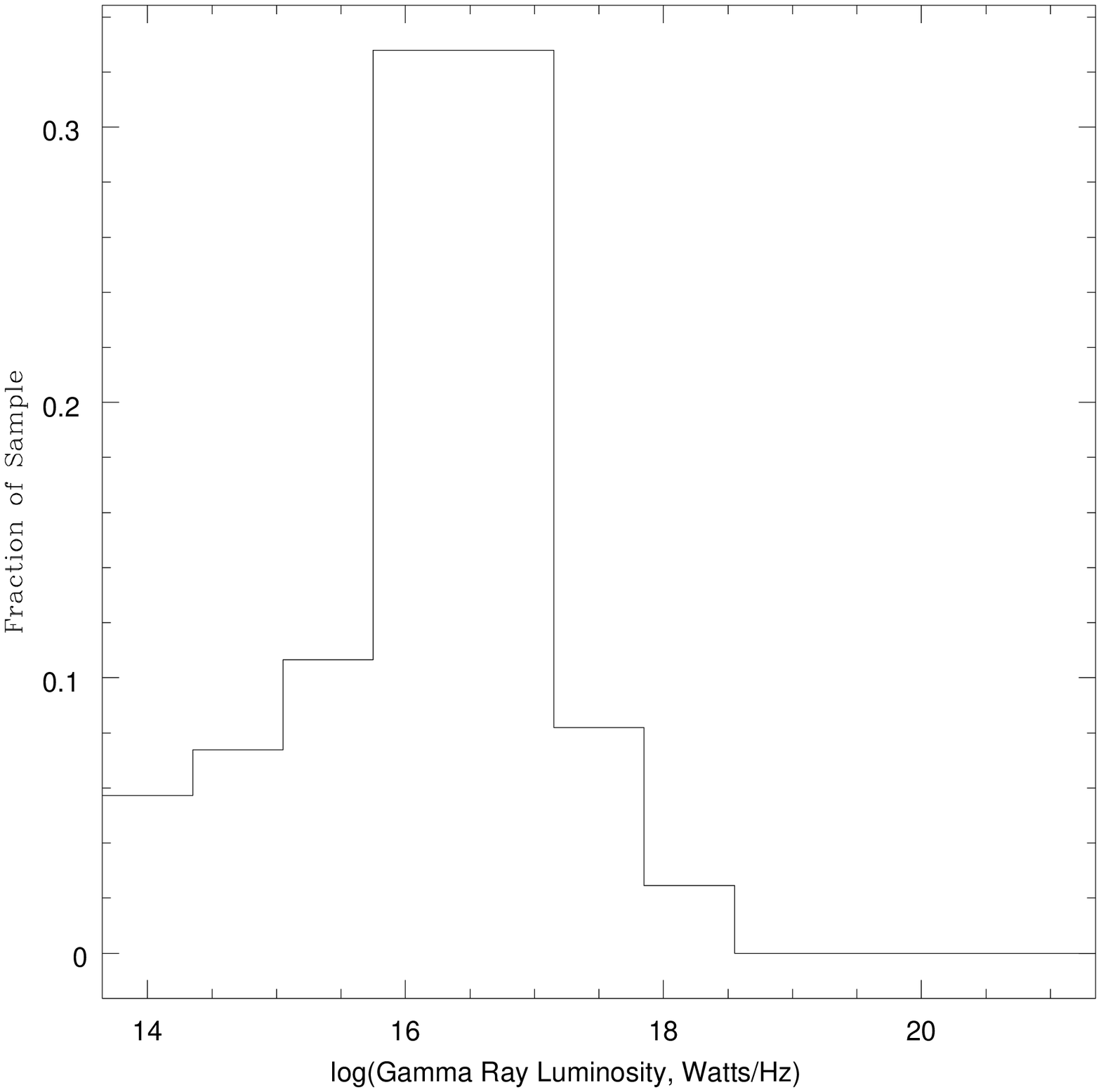}
\caption{Simulated (left) and Observed (right) Gamma Ray Luminosity
  Distributions (SSC model assumed for simulations)}
\end{figure}
\begin{figure}
\figurenum{4}
\plottwo{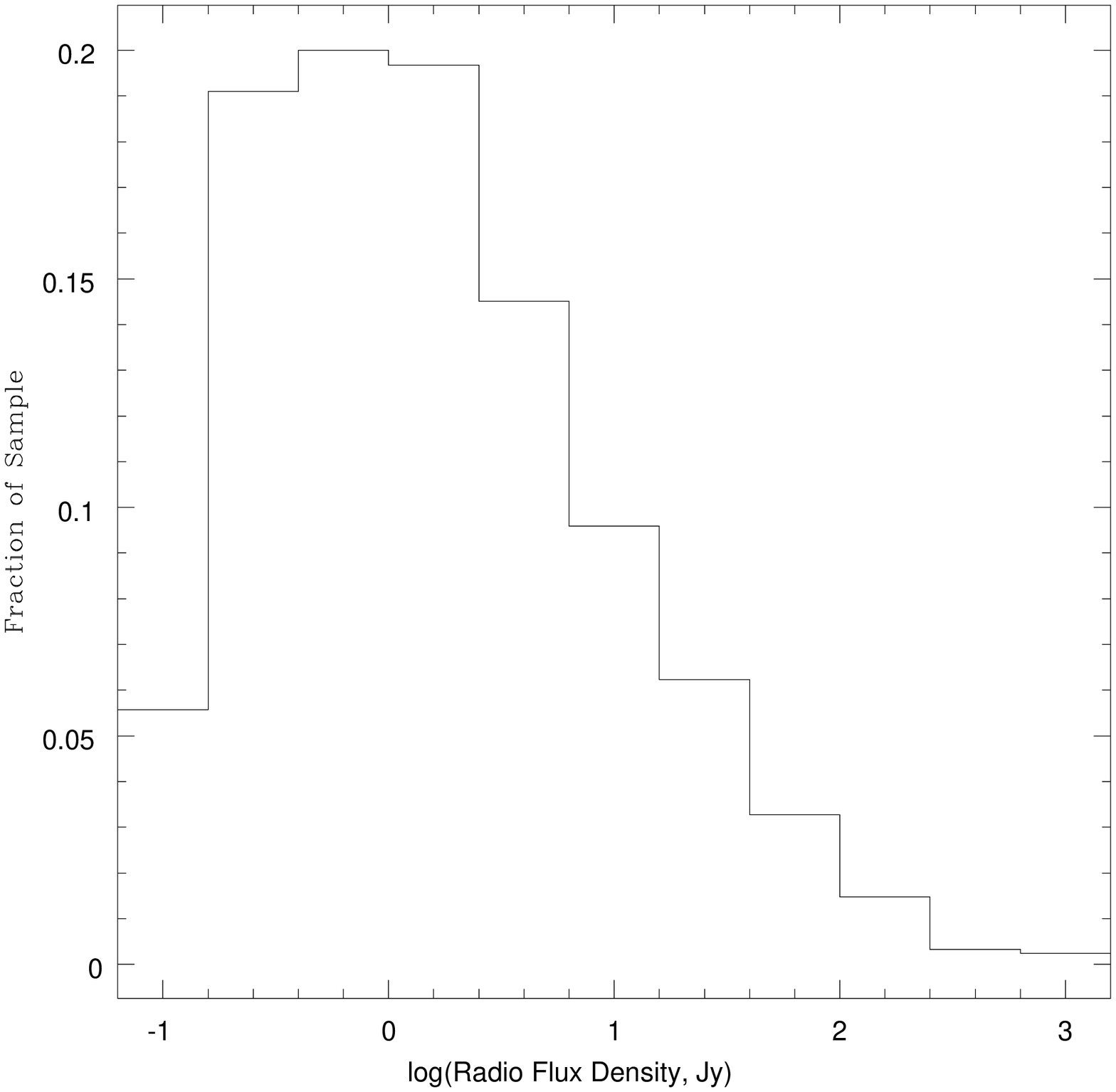}{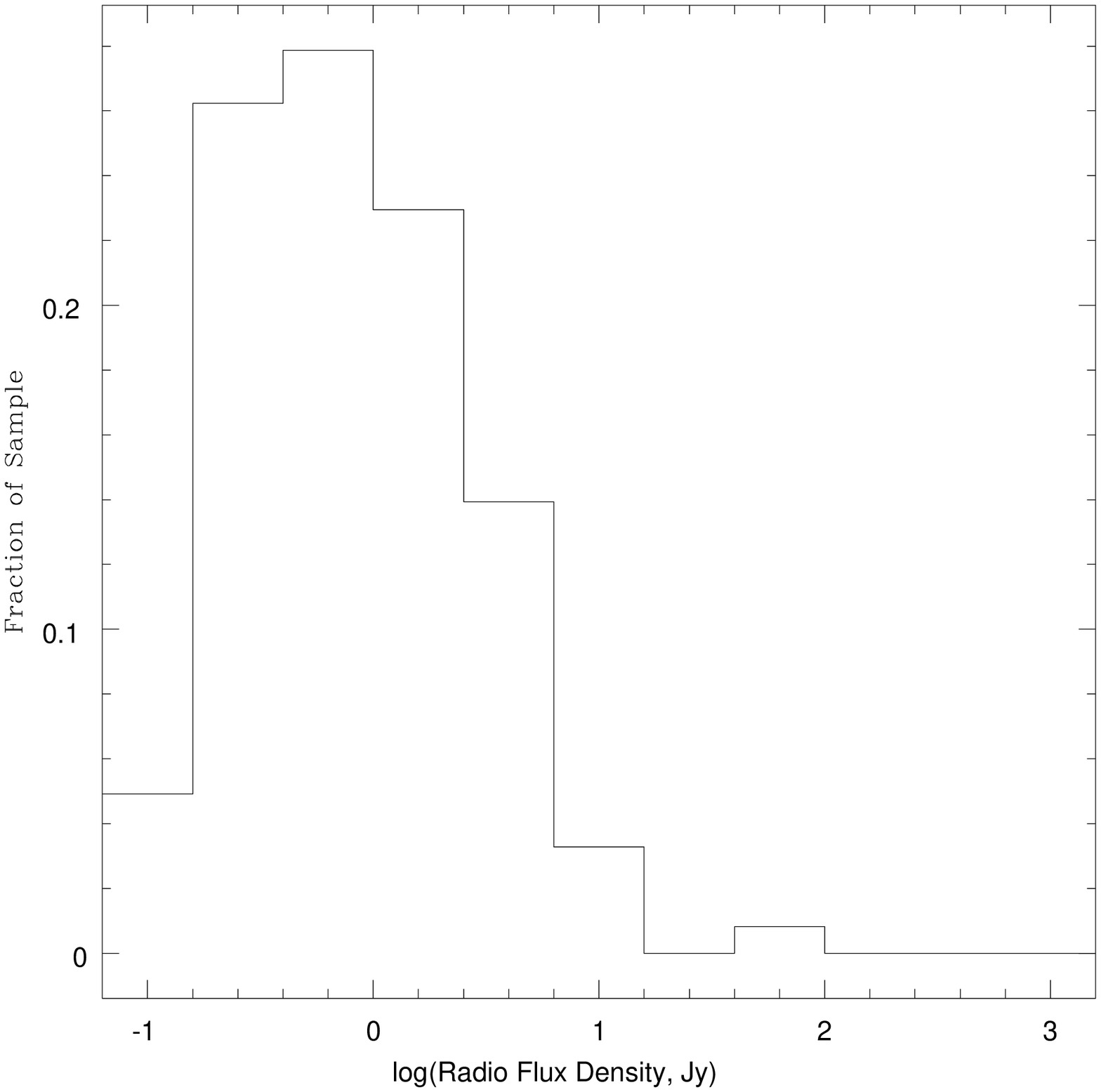}
\caption{Simulated (left) and Observed (right) Radio Flux Density
  Distributions (SSC model is assumed for simulations)}
\end{figure}
\begin{figure}
\figurenum{5}
\plottwo{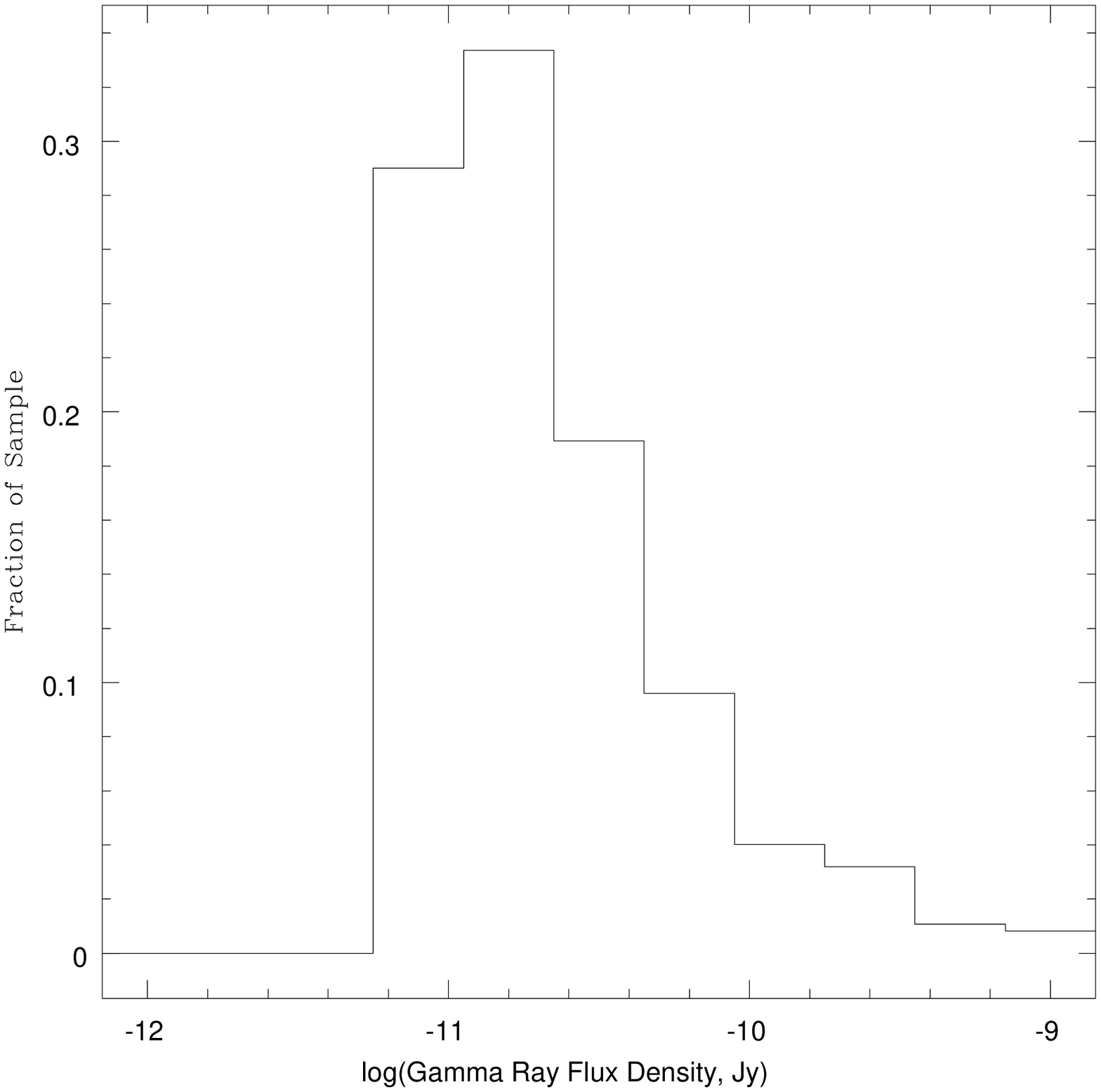}{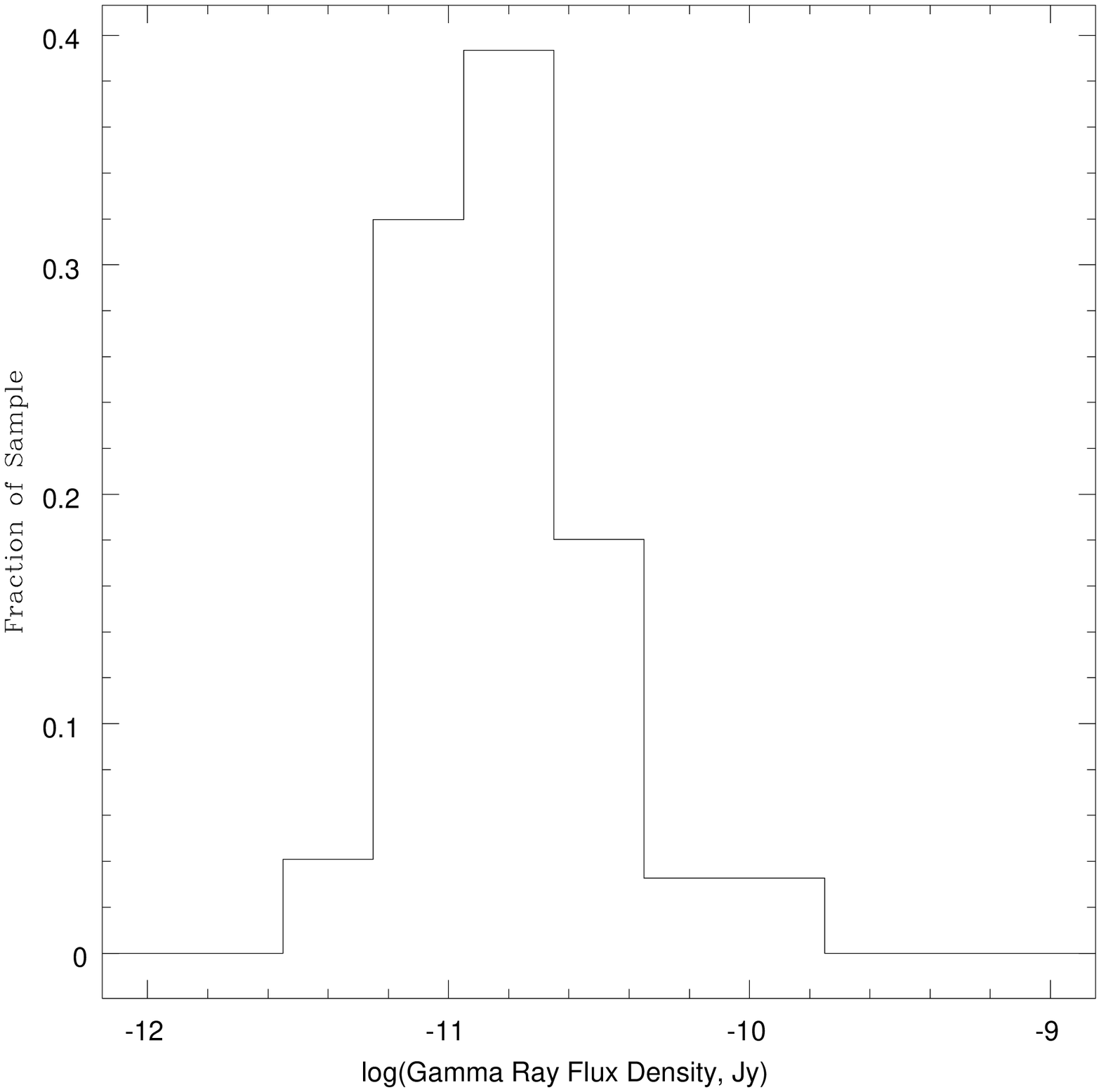}
\caption{Simulated (left) and Observed (right) Gamma Ray Flux Density
  Distributions (SSC model is assumed for the simulations)}
\end{figure}
\begin{figure}
\figurenum{6}
\plottwo{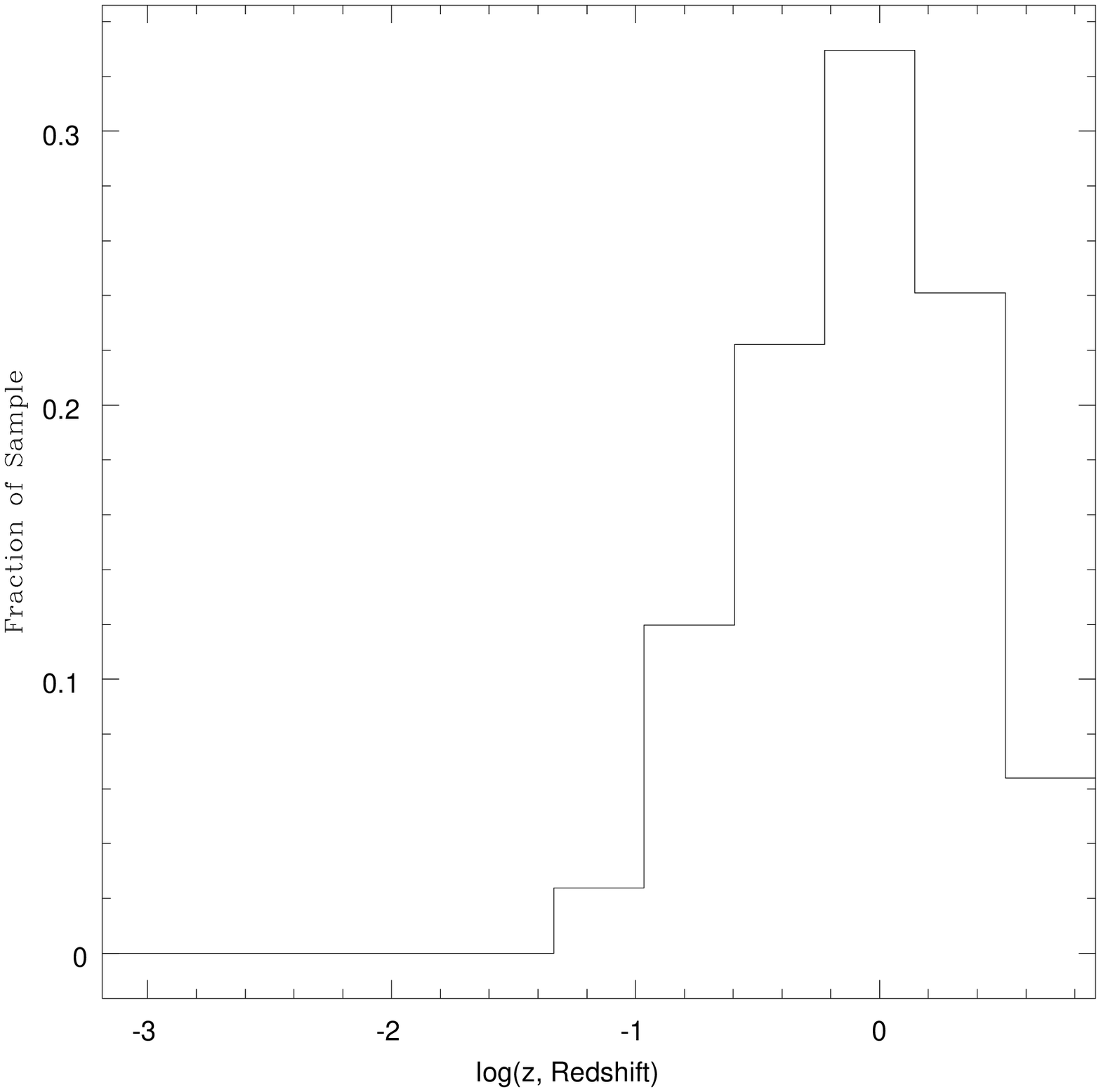}{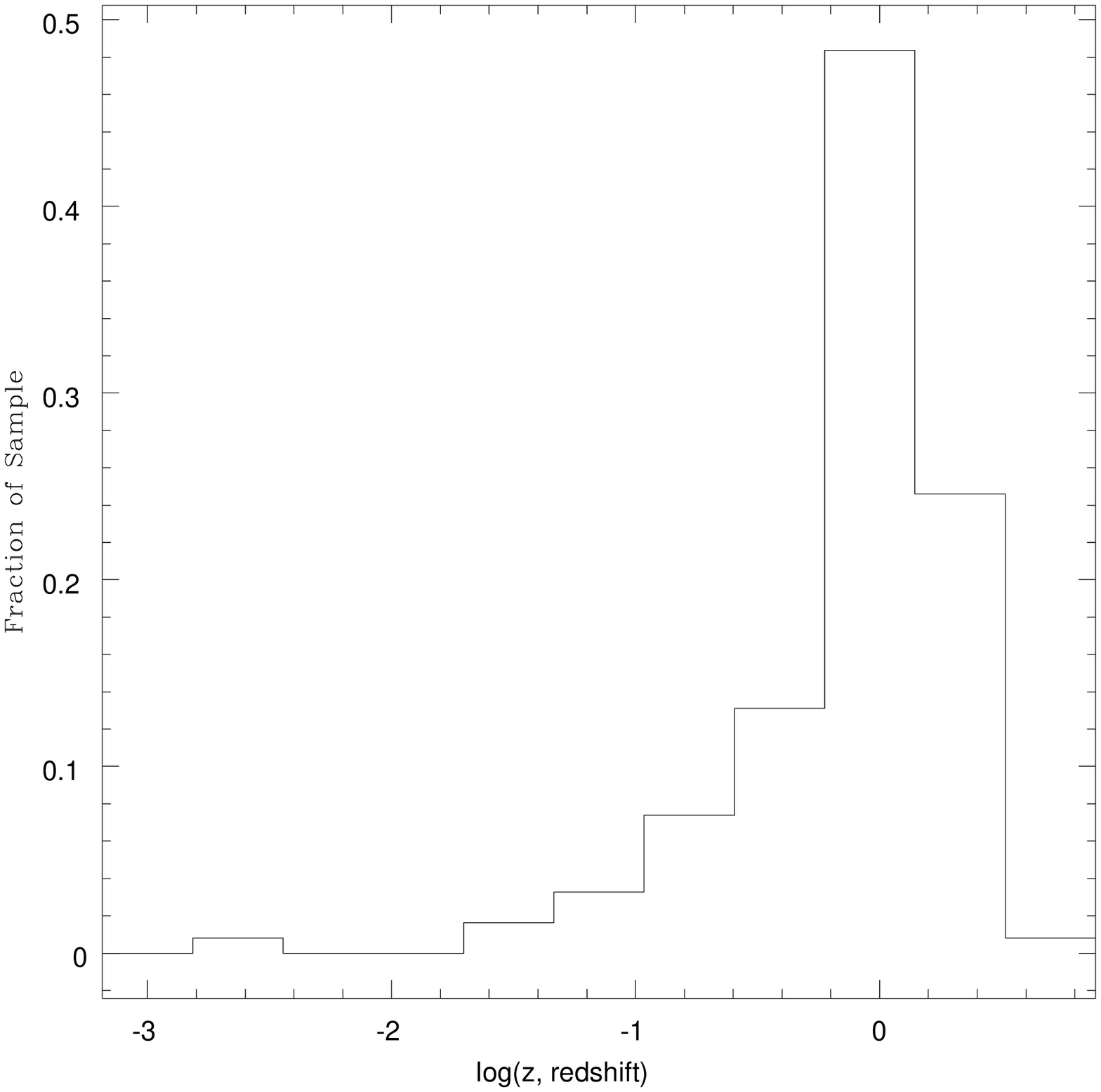}
\caption{Simulated (left) and Observed (right) Redshift Distributions
  (SSC model is assumed for the simulations)}
\end{figure}
\begin{figure}
\figurenum{7}
\plotone{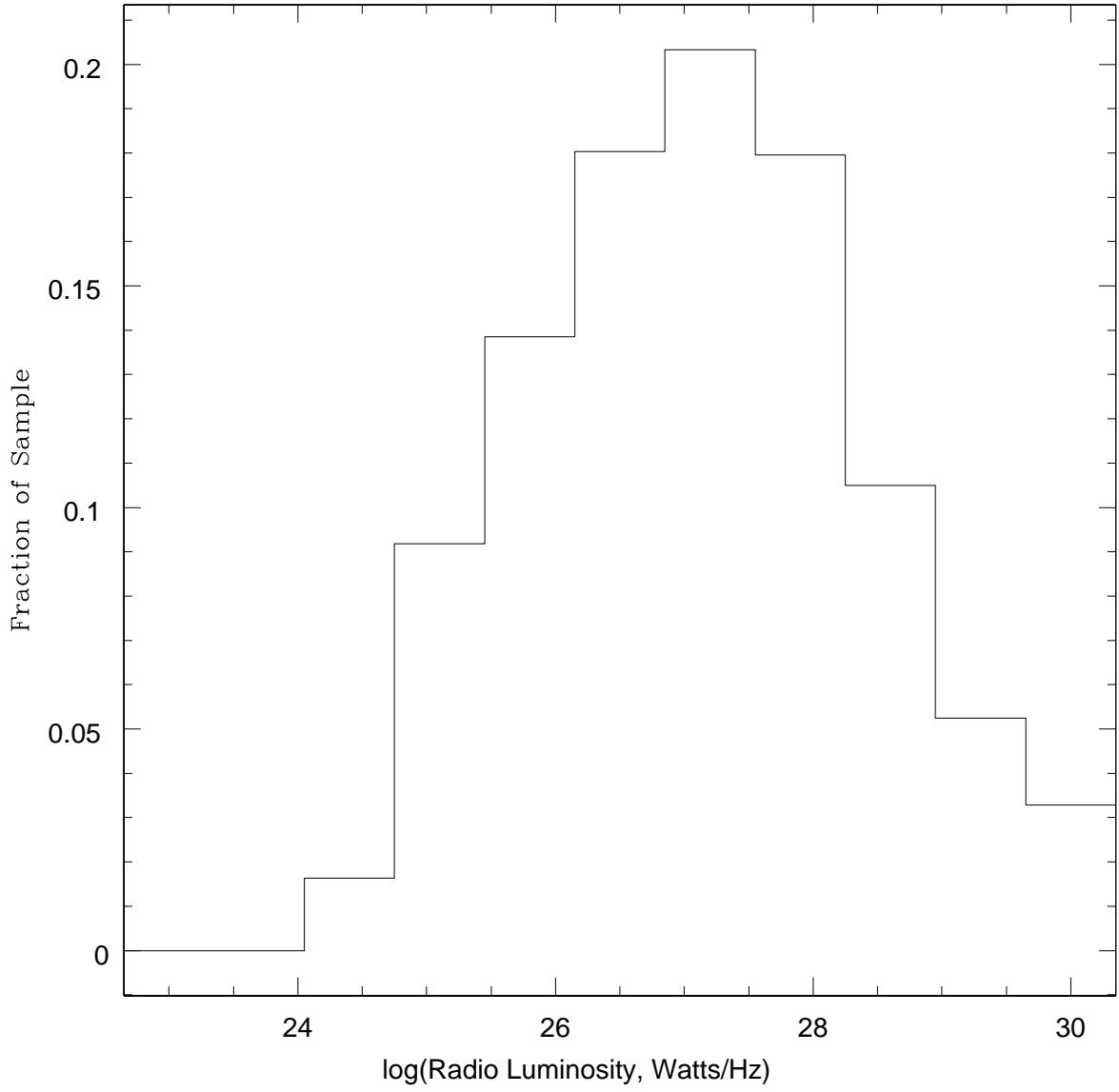}
\caption{Simulated Radio Luminosity Distribution (ECS model is assumed)}
\end{figure}
\begin{figure}
\figurenum{8}
\plotone{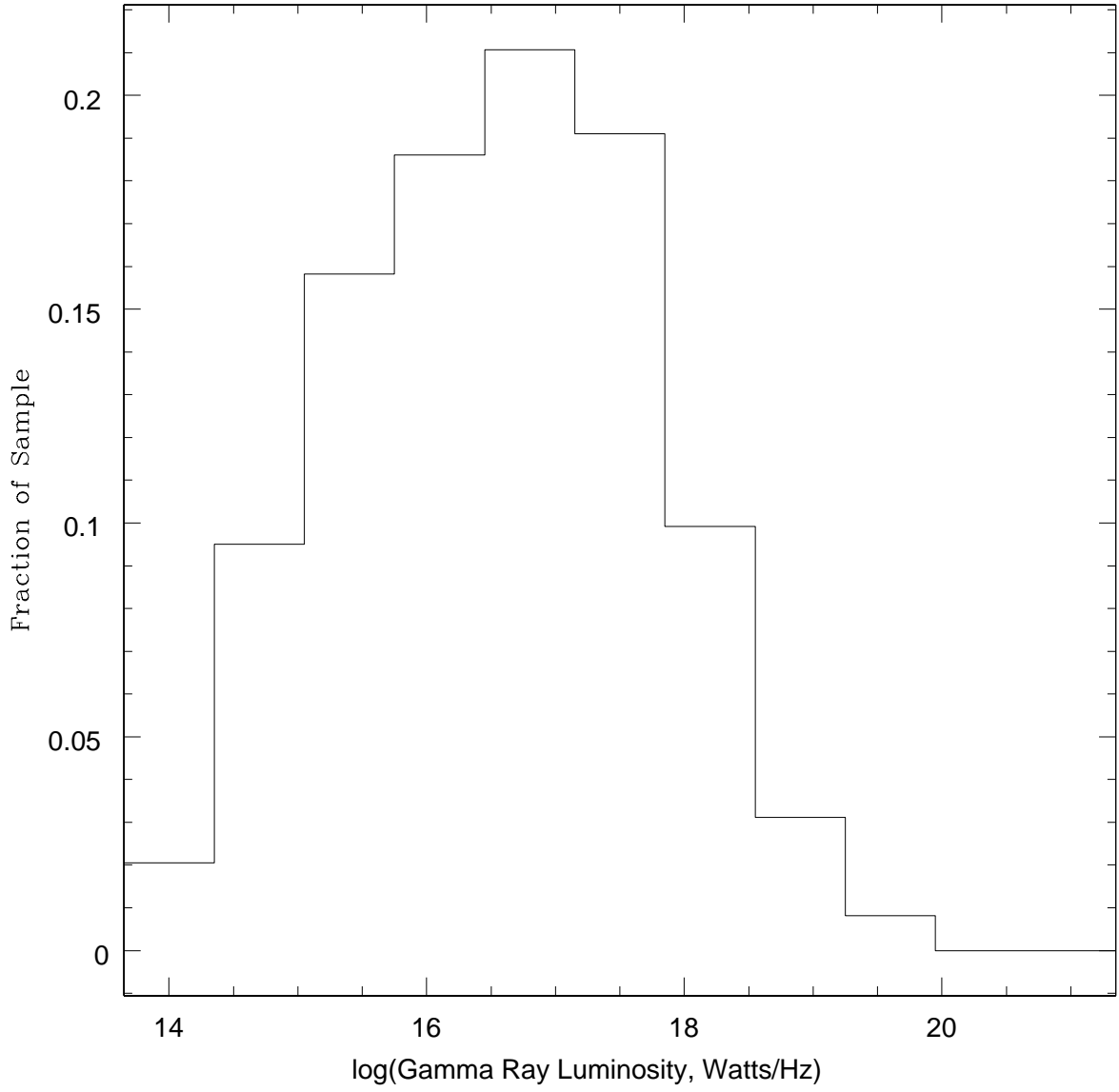}
\caption{Simulated Gamma Ray Luminosity Distribution (ECS model is assumed)}
\end{figure}
\begin{figure}
\figurenum{9}
\plotone{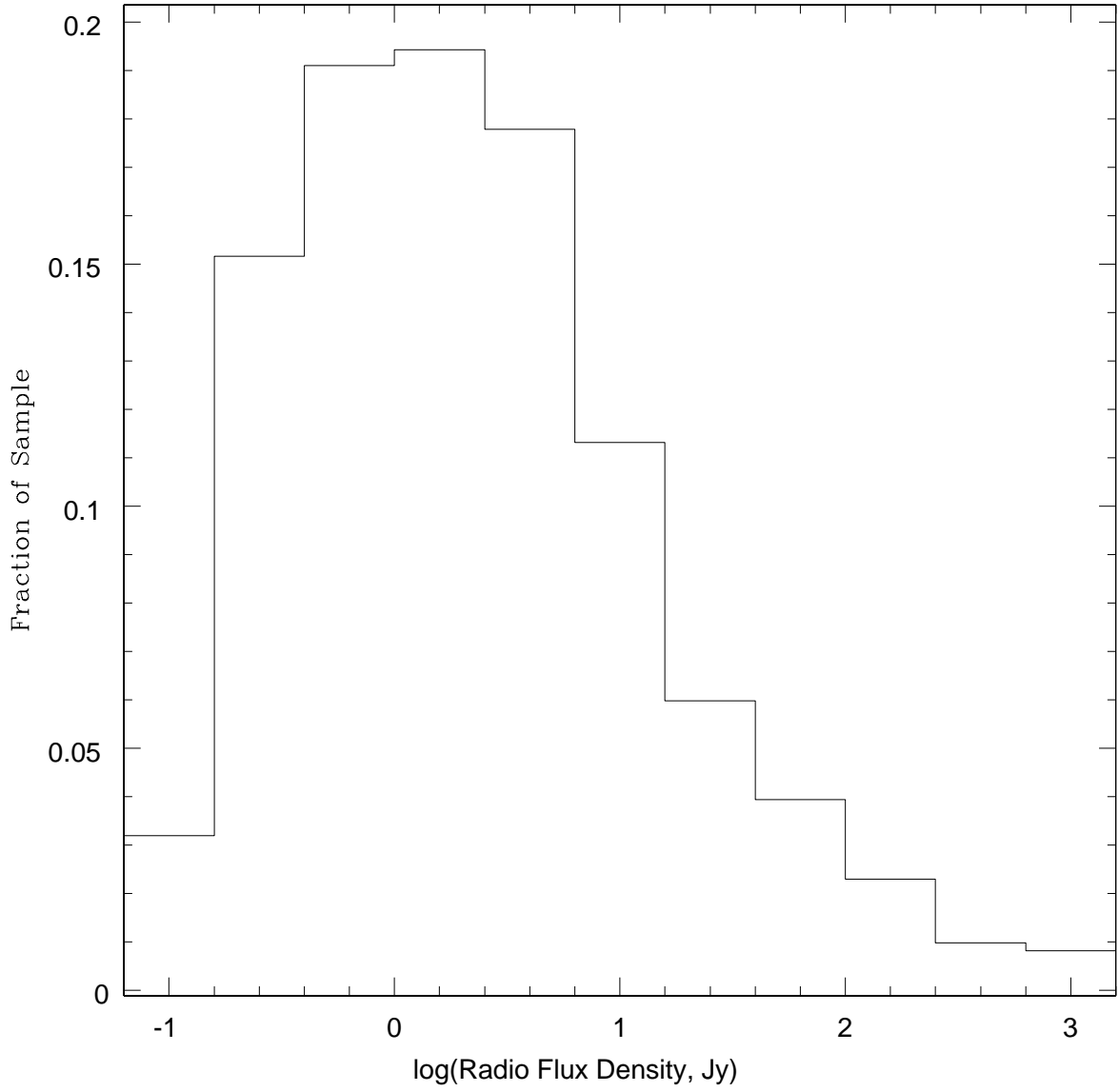}
\caption{Simulated Radio Flux Density Distribution (ECS model is assumed)}
\end{figure}
\begin{figure}
\figurenum{10}
\plotone{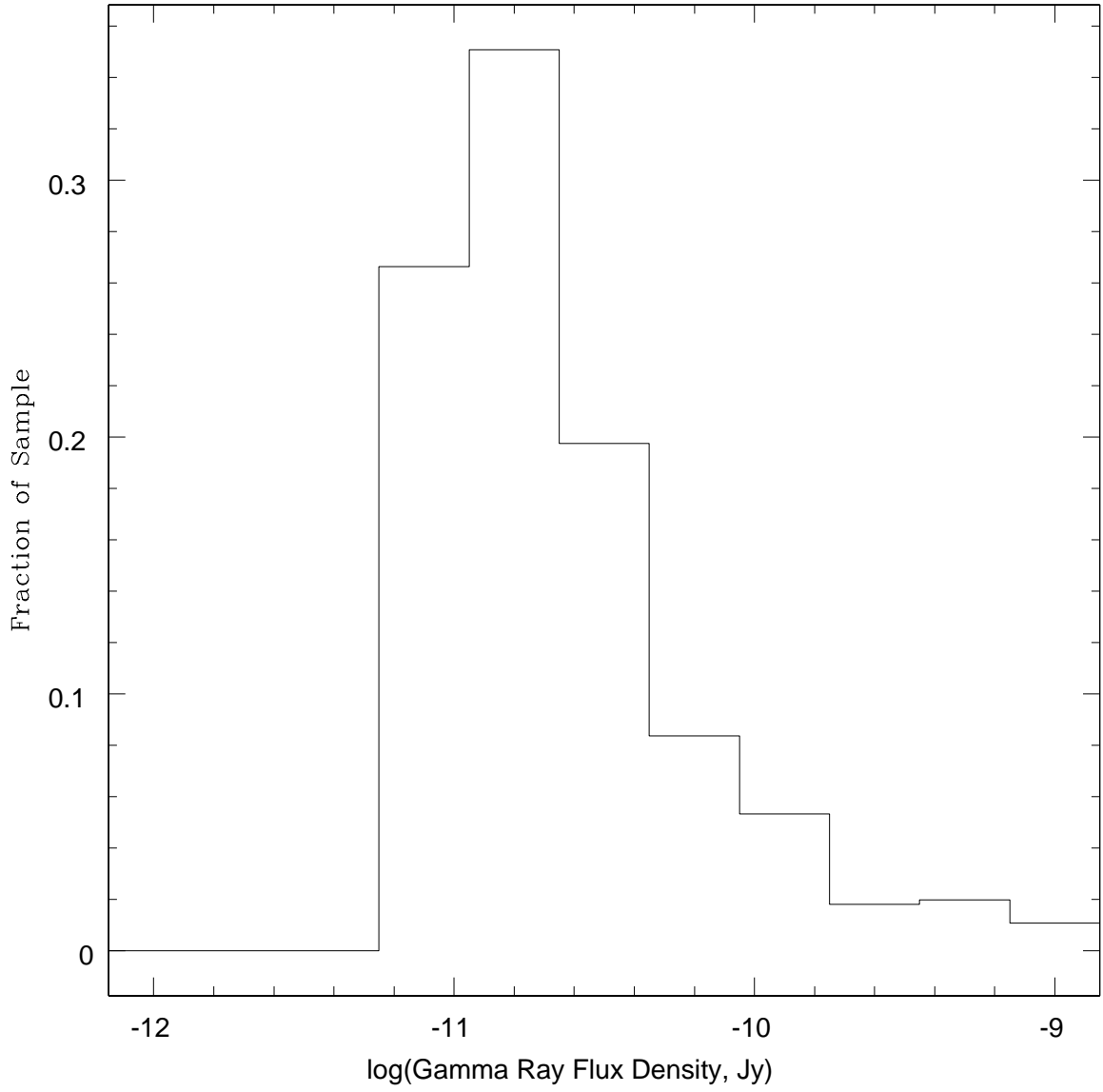}
\caption{Simulated Gamma Ray Flux Density Distribution (ECS model is assumed)}
\end{figure}
\begin{figure}
\figurenum{11}
\plotone{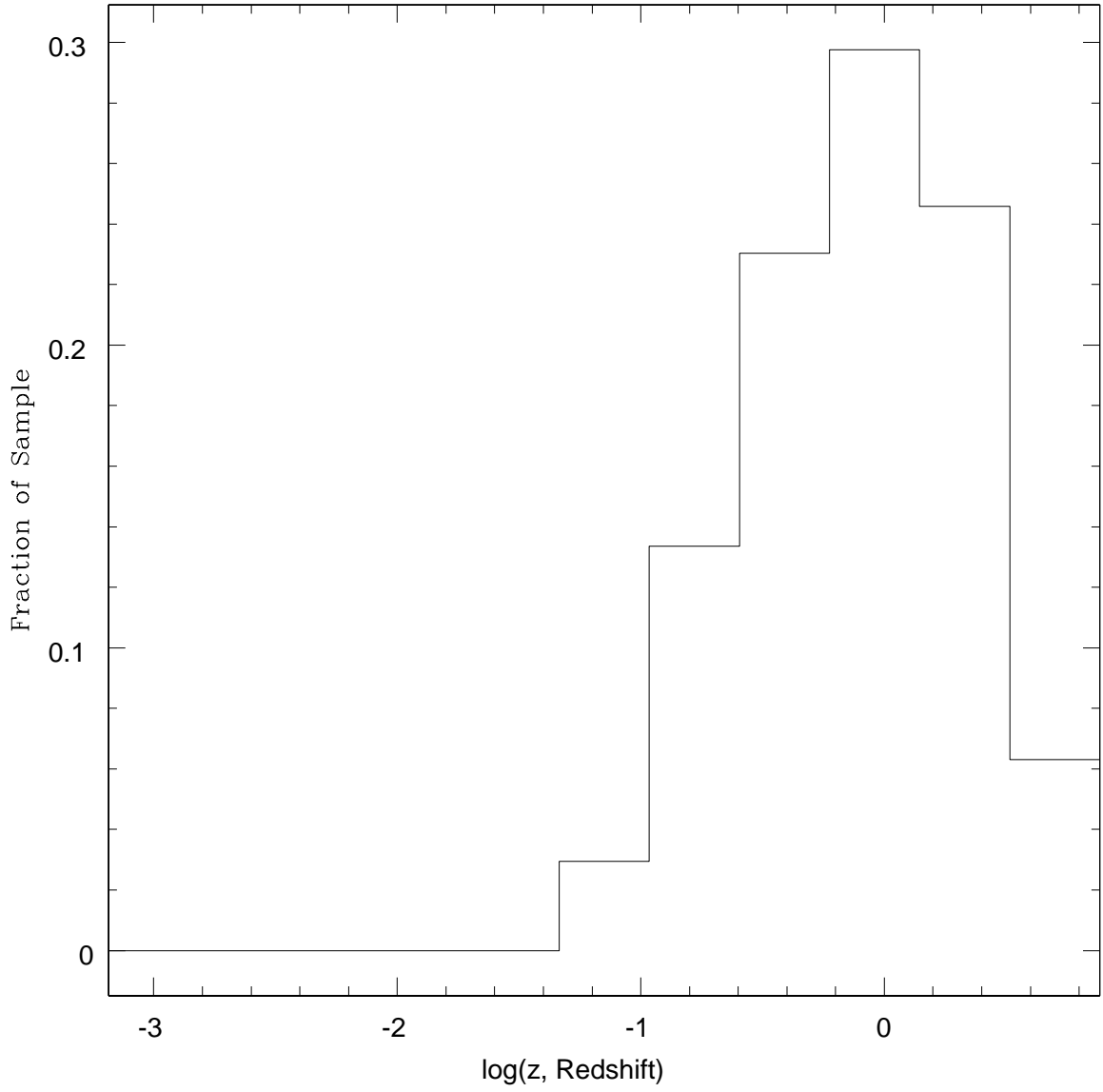}
\caption{Simulated Redshift Distribution (ECS model is assumed)}
\end{figure}
\begin{figure}
\figurenum{12}
\plottwo{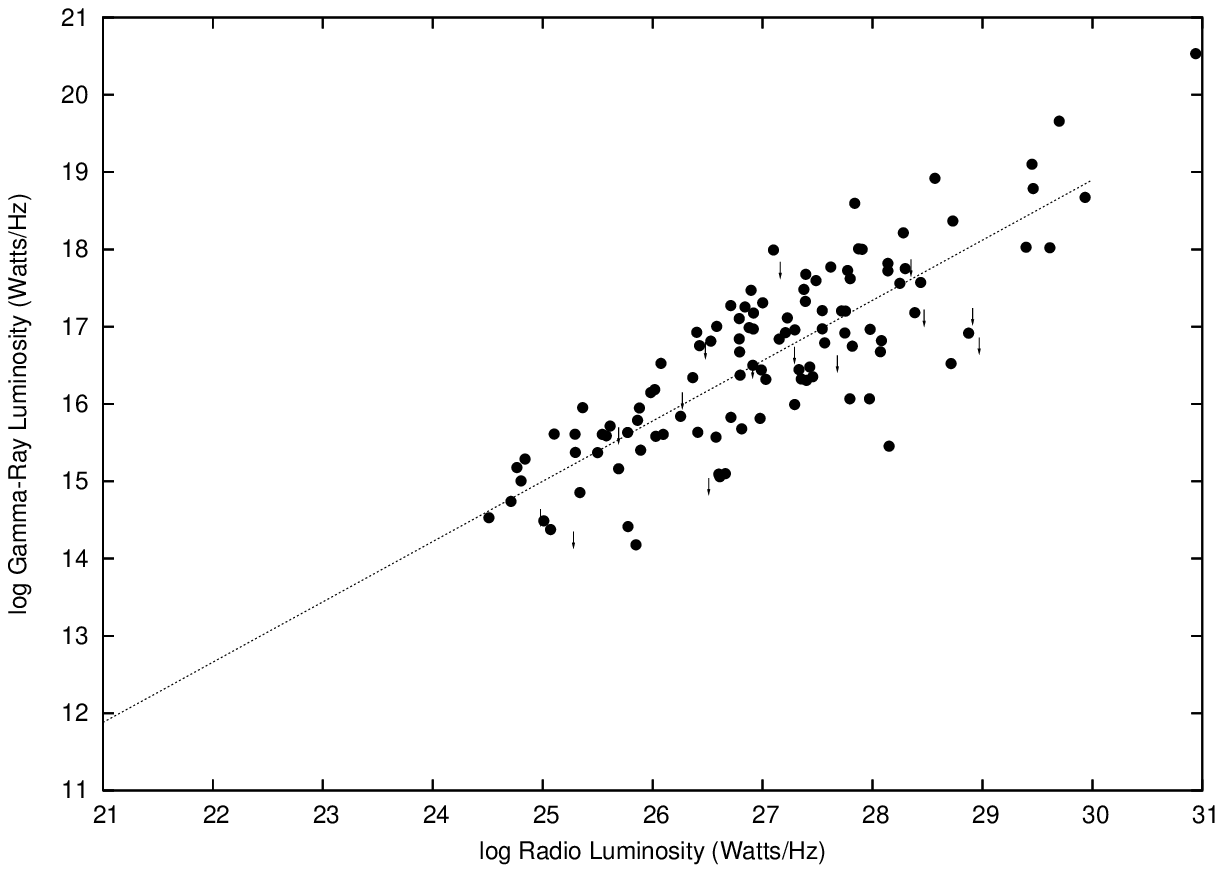}{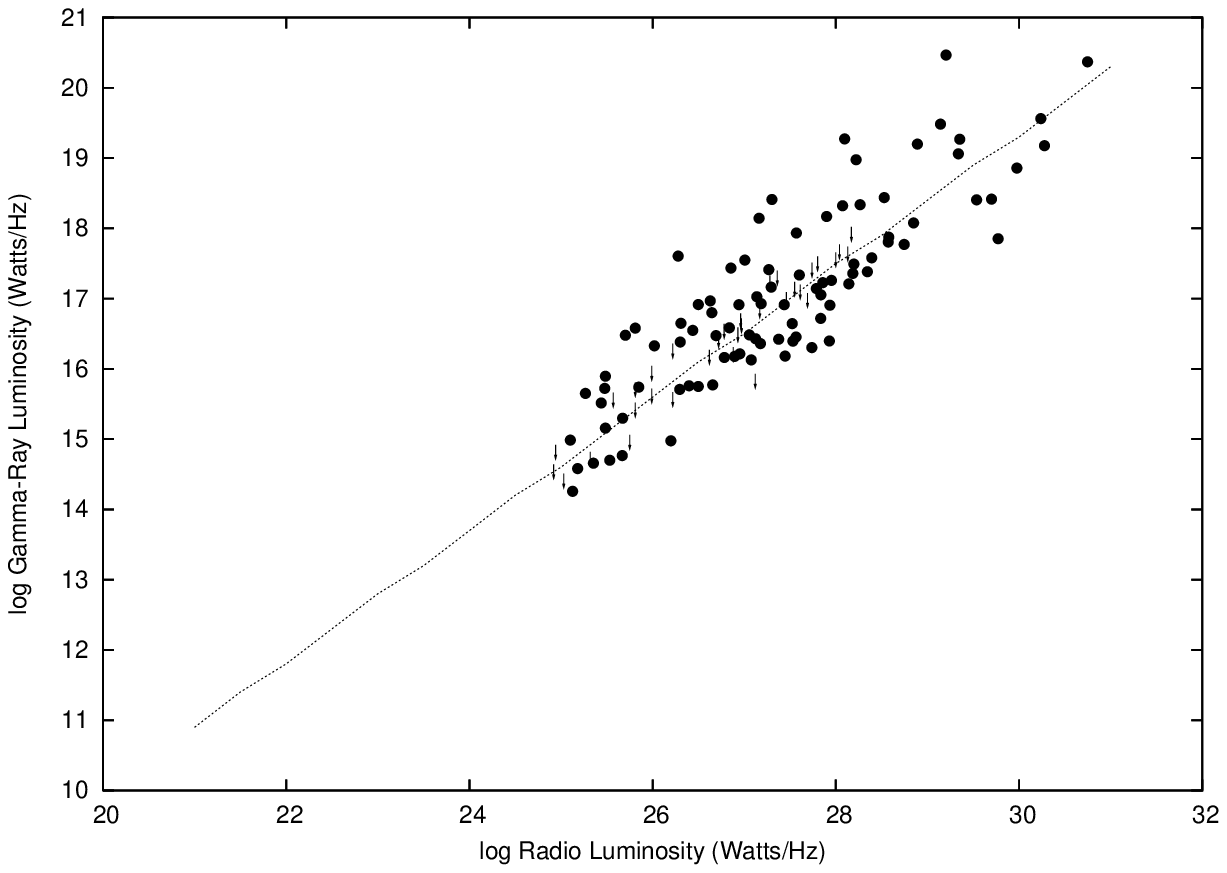}
\caption{Simulated Radio/Gamma-Ray Luminosity Correlation Diagrams  for
  SSC (left) and  ECS (right). The arrows indicate upper limits and
  the dashed line indicates the regression fit.}
\end{figure}

\clearpage

\begin{deluxetable}{lllccccccccc}
\tabletypesize{\tiny}
\tablecolumns{12}
\tablewidth{0pt}
\tablenum{1}
\tablecaption{Multiwaveband Data for EGRET Blazars}
\tablehead{\colhead {EGRET Source}& \colhead {Radio Source}&\colhead
  {z}& \colhead {log $L_r$}&\colhead {log $L_o$} &\colhead{log
    $L_g$}&\colhead{$\alpha_r$}&\colhead{$\alpha_g$
  }&\colhead{$\alpha_{ro}$}&\colhead {$\alpha_{og}$}&\colhead
  {$\delta_{var}$} &\colhead {Ref}\\
\colhead {} &\colhead {} &\colhead {} &\colhead {($Watts \, Hz^{-1}$)} &\colhead {($Watts \,Hz^{-1}$)} &\colhead
{($Watts \,Hz^{-1}$)}&\colhead {}&\colhead {}&\colhead{} &\colhead{} &\colhead{} &\colhead{}\\
\colhead{(1)} &\colhead{(2)} &\colhead{(3)} &\colhead{(4)}
&\colhead{(5)} &\colhead{(6)}&\colhead{(7)}& \colhead {(8)} &
\colhead{(9)}& \colhead {(10)} &\colhead {(11)}& \colhead {(12)}\\}
\startdata
3EG J0038-0949& J0039-0942  &2.101&28.030&24.41&16.96&-0.12&1.7 &0.69&0.83&0&2,1,6,3\\
3EG J0118+0248& J0113+0222 & 0.047& 24.26& 23.38 &13.15 & -0.1 & 1.63 & 0.18& 1.24 & 1.18&1,1,1,3\\
3EG J0130-1758& J0132-1654& 1.022& 27.74 &23.80& 16.44& -0.08& 1.5&
0.78& 0.86&0& 2,2,4,3\\
3EG J0204+1458& J0204+1514& 0.405 &27.11& 21.29& 15.53& 0.09& 1.23&
1.20&0.68&1.29& 7,1,5,3\\
3EG J0210-5055& J0210-5101& 0.999 &28.25& 24.39& 17.57& -0.07& 0.99&
0.78& 0.81& 0.31& 14,15,4,3\\
3EG J0215+1123& J0213+1213&0.252& 25.31 & 21.57 & $<$ 15.16 &-0.1& 1.03& 0.77& 0.77& 1.27&1,1,1,3\\  
3EG J0222+4253& J0222+4302 & 0.444& 26.54 & 24.11& 16.04& 0.17& 1.01& 0.50& 0.97&0 &4,1,4,3\\
3EG J0239+2815& J0237+2848&1.213& 28.48& 23.83& 16.65  &-0.1 &
1.53&0.93&0.83 &0& 4,1,4,3 \\
3EG J0237+1635& J0238+1636&0.94 & 28.51& 23.48& 17.05  & -0.5&
0.85&0.98&0.77& 0.89& 4,1,4,3  \\
3EG J0245+1758& J0242+1742&0.551& 26.38& 22.22& 15.66 &-0.1 &
1.61&0.84&0.77& 1.14 & 1,1,1,3 \\
3EG J0329+2149& J0325+2224&2.066& 28.39& 24.24& 16.82 & 0 & 1.61&
0.81& 0.83& 0  & 1,1,1,3\\
3EG J0340-0201& J0339-0146&0.852& 28.02& 23.61& 16.71 & -0.13&
0.84&0.88&0.72 & 0& 4,1,4,3 \\
3EG J0404+0700& J0407+0742& 1.133& 27.75& 24.29 & 16.42 & -0.3& 1.65&
0.67& 0.91& 0.39& 1,1,1,3 \\
3EG J0412-1853& J0416-1851& 1.536& 28.07& 24.27& $<$ 16.14& 0.23& 2.25& 0.77& 0.90& 0&4,1,4,3  \\
3EG J0422-0102& J0423-0120& 0.915& 28.28& 24.95& 16.51& -0.2& 1.44&
0.65& 0.99& 0 & 4,1,4,3\\
3EG J0423+1707&J0422+1741& 0.908&26.95&23.14&16.50&-0.5&1.43&0.73&0.77&0.42& 1,1,1,3  \\
3EG J0433+2908&J0433+2905&\nodata & 27.34&23.43& 17.03& 0&0.9& 0.78&
0.76& 0.40 &\nodata,1,3\\
3EG J0442-0033&J0442-0017& 0.844&27.32& 23.27& 16.36& 0.39& 1.37&
0.84&0.81 &1.59& 4,1,4,3\\
3EG J0450+1104&J0449+1121&1.207&28.11& 23.44& 16.85&-0.2& 1.27& 1.06&
0.69&1.13 & 5,1,5,3\\
3EG J0456-2338&J0457-2324& 1.003& 27.98& 23.66& 15.84&0&2.14& 0.84&
0.9&0.21& 5,1,5,3\\
3EG J0458-4635&J0455-4615&0.858 &27.66& 24.02& 15.94&0.18&1.75&0.74&
0.94&0.41 & 4,15,4,3\\
3EG J0459+0544&J0502+0609&1.106 &27.48& 23.40& 16.32& 0.28&1.36& 0.82&
0.83&0.74& 5,1,5,3\\
3EG J0459+3352&J0503+3403&0.149 &25.16& 22.02 & 14.64&0.38&1.54&0.65&
0.89&0.59& 1,1,1,3\\
3EG J0500-0159&J0501-0159&2.2286& 29.53& 24.76& 17.20&-0.25&1.45&0.91&
0.86&1.09& 4,2,4,3 \\
3EGJ0510+5545&J0514+5602&2.19&28.02&23.50&17.66&0.18&1.19&0.91&0.67&0&1,1,1,3\\
3EG J0512-6150&J0506-6109& 1.093&27.90& 24.52 & 16.36&0.21& 1.40&
0.68&0.95&0&4,15,4,1 \\
3EG J0530-3626&J0529-3555&\nodata&27.29&24.08&16.47&-0.42&1.63&0.64&1.04&0.58& \nodata,2,7,3\\
3EGJ0530+1323&J0530+1331&2.07&29.31&24.03&18.04&-0.3&1.46&1.01&0.55&0.74&
5,1,5,3\\
3EG J0531-2940&J0539-2839&3.104&29.07 &24.52&17.27&0.06&1.47&0.89&0.81&0.85& 4,2,4,3\\
3EG J0533+4751&J0533+4822&1.16&27.68&23.67&16.61&-0.1&1.55&0.75&0.84&
0.0& 1,1,1,3\\
3EGJ0540-4402&J0538-4405&0.896&28.56&25.20&17.08&-0.17&1.41&0.66&0.95&0.75&4,15,4,3\\
3EGJ0542-0655&J0541-0541&0.839&27.49&22.95&$<$16.45&-0.06&1.0&0.91&0.77&1.38&5,2,5,3\\
3EGJ0542+2610&J0540+2507&0.62&26.40&24.56&15.96&0.12&1.67&0.90&0.70&0.57&
1,1,1,3\\ 
3EG J0721+7120&J0721+1721&0.3&26.02&23.61&15.56&0.09&1.19&0.49&0.97&0&4,1,4,3\\
3EG J0737+1721&J0738+1742&0.424&27.14&24.24&15.69&-0.1&1.6&0.58&1.02&0&4,1,4,3\\
3EG J0743+5447&J0742+5444&0.723&26.36&24.84&16.32&0.36&1.03&0.55&0.88&1.21&5,1,5,3\\
3EG J0808+5114&J0807+5117&1.14&27.57&23.95&16.24&-0.4&1.76&0.69&0.88&0&5,1,5,3\\
3EG J0828+0508&J0831+0429&0.1736&25.77&22.65&14.88&0&1.47&0.64&0.93&0&4,1,4,3\\
3EG J0829+2413&J0830+2410&0.939&27.48&24.18&16.73&0&1.42&0.66&0.87&\nodata&4,1,4,3\\
3EG J0845+7049&J0841+7053&2.172&28.77&25.49&16.99&0.42&1.62&0.67&0.96&0.62&4,1,4,3\\
3EG J0852-1216&J0850-1213&0.566&26.87&23.77&16.31&-0.37&0.58&0.61&0.90&1.21&5,2,5,3\\
3EG J0853+1941&J0854+2006&0.306&26.80&24.23&15.41&-0.3&1.03&0.51&1.06&0&4,1,4,3\\
3EG J0917+4427&J0920+4441&2.18&28.93&24.49&17.47&-0.1&1.19&0.86&0.81&0.0&4,1,4,3\\
3EG J0952+5501&J0957+5522&0.901&27.64&23.87&16.42&0.39&1.12&0.78&0.88&0.39&4,1,4,3\\
3EG J0958+6533&J0958+6533&0.368&26.64&24.70&15.33&-0.3&1.08&0.38&0.88&0.96&4,1,4,3\\
3EG J1052+5718&J1058+5628&0.14&25.88&23.54&15.18&0.09&1.51&0.31&1.09&0.24&5,1,5,3\\
3EG J1104+3809&J1104+3812&0.031&23.85&22.32&13.46&0&0.57&0.32&1.07&0.35&4,1,4,3\\
3EG J1133+0033&J1133+0040&1.633&27.84&24.07&16.21&0.05&1.73&0.74&0.89&0.82&1,1,1,3\\
3EG J1200+2847&J1159+2914&0.729&27.34&23.99&16.18&0.24&0.98&0.68&0.93&1.17&4,1,4,3\\
3EG J1219-1520&J1222-1645&\nodata&26.87&23.42&15.95&0.03&1.52&0.65&0.89&1.10&\nodata,2,8,3\\
3EG J1222+2841&1221+2813&0.102&25.25&22.13&14.45&-0.2&0.73&0.64&0.93&0.62&4,1,4,3\\
3EG J1224+2118&J1224+2122&0.435&26.66&23.20&15.79&0.34&1.28&0.71&0.88&0.45&4,1,4,3\\
3EG J1227+4302&J1224+4335&1.872&27.69&23.74&$<$16.97&0.26&1.5&0.80&0.79&1.36&1,1,1,3\\
3EG J1229+0210&J1229+0203&0.158&27.20&24.02&14.74&0.04&1.58&0.65&1.12&0.46&4,1,4,3\\
3EG J1236+0457&J1231+0418 &1.03&27.21&23.99&16.21&0.05 &1.48 &0.64&0.91&0.53&1,1,1,3\\
3EG J1246-0651&J1246-0730&1.286&27.86&24.26&16.42&-0.08&1.73&0.71&0.6&\nodata&4,2,4,3\\
3EG J1255-0549&J1256-0547&0.538&28.16&23.34&16.86&-0.2 &0.96&0.97&0.78&0.9&4,2,4,3\\
3EG J1310-0517&J1312-04242&0.824&26.96&23.24&16.15&-0.15&1.34&0.76&0.82&\nodata&1,1,1,3\\
3EG J1323+2200& J1327+2210&1.40&28.66&24.06&16.79&-0.5 &0.86&0.74&0.95&1.09&5,1,5,3\\
3EG J1324-4314& J1325-4301&0.0018&21.78&22.00&10.74&0.66 &1.58&-0.05&1.36&0&16,7,13,3\\
3EG J1329+1708& J1333+1649&2.09&29.71&25.36&16.76&-0.1 &1.41&0.57&0.99&0.62&4,1,4,3\\
3EG J1339-1419 &J1337-1257&0.539&27.56&23.56&15.67&-0.21&1.62&0.80&0.93&0.81&4,2,4,3\\
3EG J1347+2932 &J1343+2844&0.91&26.81&24.24&16.24&0.13&1.51&0.51&0.93&0.64&9,1,9,3\\
3EG J1409-0745 &J1408-0752&1.494&28.03&24.28&17.32&0&1.29&0.74&0.81&1.42&4,2,4,3\\
3EG J1424+3734 &J1419+3821&1.83&28.47&24.08&16.35&-0.1&2.25&0.85&0.84&\nodata&4,1,4,3\\
3EG J1429-4217& J1427-4206&1.522&28.59&24.16&17.08&0.15&1.13&0.80&0.88&0.93&4,1,4,3 \\
3EG J1457-1903 &J1459-1810&\nodata &26.98&$<$24.60&16.15&-0.24&1.67&0.78&0.79&0.49&\nodata,2,7,3\\
3EG J1500-3509&J1457-3539&1.422& 27.94&23.97&16.36&0.04&1.99&0.79&0.85&0&2,2,2,3\\ 
3EG J1504-1537&J1502-1508&\nodata&26.88&$<$23.02&$<$16.58&-0.18&\nodata&0.76&0.76&1.24&\nodata,2,\nodata,3\\
3EG J1512-0849&J1512-0905&0.360&26.78&23.39&15.63&0.12&1.47&0.69&0.92&0&4,2,4,3\\
3EG J1517-2538&J1517-2422&0.049&24.75 & 19.84 &13.39& 0.02&1.66&1.02&0.75&0&2,2,2,3\\
3EG J1527-2358&J1532-2310&2.289&28.06&24.11&$<$16.81&-0.09&1.67&0.76&0.81&0.92&2,2,2,3\\
3EG J1605+1553&J1603+1554&0.11&24.66&23.77&14.50&-0.5&1.47&0.17&1.12&0&1,1,1,3\\
3EG J1608+1055&1608+1029&1.23&28.27&23.59&16.86&-0.1&1.06&0.84&0.82&1.09&5,1,5,3\\
3EG J1607-1101&J1612-1133&\nodata&27.11&$<$23.02&$<$16.36&-0.58&\nodata&0.78&0.79&1.45&\nodata,2,\nodata,3\\
3EG J1612-2618&J1611-2612&\nodata&26.80&$<$23.02&16.10&-0.17&1.71&0.75&0.79&1.04&\nodata,2,\nodata,3\\
3EG J1614+3424&J1613+3412&1.40&28.58&24.56&17.15&0.13&1.42&0.81&0.86&0&4,1,4,3\\
3EG J1621+8203&J1632+8232& 0.02&23.53&21.93&12.67&0.05& 1.29 &0.33&1.11&0&5,2,7,3\\
3EG J1625-2955&J1626-2951&0.815&27.75&23.18&17.06&0.00&1.07&1.02&0.67&1.62&5,2,5,3\\
3EG J1626-2519&1625-2527&0.786&27.74&22.63&16.60&0.45&1.21&1.17&0.65&1.05&5,2,5,3\\
3EG J1634-1434&J1628-1415&1.025&27.15&23.08&16.65&0.14 &1.15&0.82&0.76&0&4,2,4,3\\
3EG J1635-1751&J1629-1720&\nodata&27.27&$<$22.96&$<$16.36&-0.56&\nodata&0.83&0.78&0.83&\nodata,2,\nodata,3\\
3EG J1635+3813&J1635+3808&1.81&28.86&24.79&17.94&0.04&1.15&0.80&0.79&0&4,1,4,3\\
3EG J1646-0704&J1644-0743 &0.139&24.58&20.44&14.59&-0.08&1.39 &0.85&0.70&0.75&2,2,2,3\\
3EG J1718-3313&J1717-3342&\nodata&27.56&$<$22.96&16.57&-0.06&1.59&0.92&0.74&0.82&\nodata,2,\nodata,3\\
3EG J1720-7820&1723-7713&\nodata&27.38&23.61&16.12&-0.55&1.74&0.69&0.88&0.64&\nodata,7,8,3\\
3EG J1727+0429&J1728+0427&0.29&26.01&22.97&15.33&0.04&1.67&0.62&0.91&0&4,1,4,3\\
3EG J1733+6017&J1722+6105&2.06&28.04&24.07&16.55&-0.1&2.0&0.75&0.83&0.29&1,1,1,3\\
3EG J1733-1313&1733-1304&0.9&28.37&23.63&16.96&-0.02&1.23&0.95&0.78&0.37&4,2,4,3\\
3EG J1735-1500&J1738-1503&\nodata&27.43&23.33&$<$16.23&0.15&2.24&0.83&0.80&0.87&10,2,10,3\\
3EG J1738+5203&J1740+5211&1.38&28.33&24.14&16.97&-0.2&1.42&0.81&0.83&0.51&4,1,4,3\\
3EG J1744-0310&J1743-0350&1.05&28.16&23.78&16.52&-0.22&1.42&0.86&0.84&0.59&4,1,4,3\\
3EG J1800-3955&J1802-3940&\nodata&26.79&22.80&15.27&0.07&2.1&0.82&0.91&0&17,2,\nodata,3\\
3EG J1806-5005&J1808-5011&1.61&27.80&23.63&16.25&-0.33&1.93&0.84&0.82&0.88&12,7,12,3\\
3EG J1824+3440&J1826+3431&1.81&27.78&25.23&$<$17.16&0.25&1.03&0.51&0.95&0.01&1,1,1,3\\
3EG J1828+0142&J1826+0149&1.77&28.41&25.98&$<$16.91&-0.2&1.76&1.01&0.70&1.6&1,1,1,3\\
3EG J1832-2110&J1833-2103&2.510&29.63&23.46&17.56&0.25&1.59&1.25&0.64&0.62&5,2,5,3\\
3EG J1850-2652&J1848-2718&\nodata&27.71&$<$22.96&16.29&-1.26&1.19&0.88&0.78&0.81&\nodata,2,\nodata,3\\
3EG J1904-1124&J1905-1153&\nodata &26.81&$<$22.96&16.51&0.26&1.6&0.79&0.75&0.08&\nodata,2,\nodata,3\\
3EG J1911-2000&J1911-2006&1.119&28.16&23.73&16.77&0.11&1.39&0.89&0.81&0.3&11,2,11,3\\
3EG J1921-2015&J1923-2104&0.871&27.82&24.11&$<$16.15&0.19&0.75&0.95&1.24&\nodata&11,2,11,3\\
3EG J1935-4022&J1937-3958&0.965&27.82&23.58&16.02&-0.15&1.86&0.77&0.91&1.34&5,2,5,3\\
3EG J1937-1529&J1939-1525&1.66&28.20&24.76&$<$15.94&0.0&2.45&0.83&0.88&1.35&4,2,4,3\\
3EG J1959+6342&J2006+6424&1.57&28.41&$<$23.50&16.94&-0.3&1.45&0.95&0.75&0&5,1,5,3\\
3EG J2006-2321&J2005-2310&0.830&26.79&23.29&16.13&0.14&1.33&0.71&0.84&1.31&2,2,2,3\\
3EG J2025-0744&J2025-0735&1.388&27.88&24.15&17.07&0.33&1.38&0.76&0.82&0.87&5,2,5,3\\
3EG J2027+3429&J2025+3343&0.22&26.32&21.08&15.39&-0.4&1.28&1.07&0.68&0&1,1,1,3\\
3EG J2034-3110&J2030-3039&\nodata &27.08&25.10&15.44&-0.05&2.43&0.82&0.84&1.18&12,2,12,3\\
3EG J2036+1132&J2034+1154&0.60&26.82&23.13&15.80&0.46&1.83&0.73&0.85&0&5,1,5,3\\
3EG J2046+0933&J2049+1003&\nodata &27.84&$<$22.96&16.40&-0.6&1.22&0.94&0.77&0&\nodata,\nodata,1,3\\
3EG J2055-4716&J2056-4714&1.49&28.38&23.99&17.02&0.32&1.04&0.89&0.82&0.93&4,15,4,3\\
3EG J2100+6012&J2102+6015&4.57&28.59&26.13&18.33&0.35&1.21&0.52&0.88&0.13&2,2,2,3\\
3EG J2158-3023&J2158-3013&0.12&24.61&23.29&14.51&0.44&1.35&0.27&1.06&0.59&4,2,4,3\\
3EG J2202+4217&J2202+4216 &0.07&25.31&22.58&13.85&0.31&1.6&0.57&1.05&0.8&1,1,4,3\\
3EG J2206+6602&J2208+6519&1.12&27.14&22.69&16.98&0.33&1.29&0.91&0.66&0&1,1,1,3\\
3EG J2209+2401&J2212+2355&1.13&27.76&23.70&16.32&-0.1&1.48&0.80&0.86&0.91&1,1,1,3\\
3EG J2232+1147&J2232+1143 &1.04&28.08&24.27&16.70&0.48&1.45&0.79&0.88&0.48&4,1,4,3\\
3EG J2254+1601&J2253+1608&0.86&28.50&24.54&17.10&0.1&1.21&0.80&0.88&0.52&4,1,4,3\\
3EG J2255+1943&J2253+1942&0.28&25.75&22.53&14.94&-0.1&1.36&0.66&0.91&1.18&5,1,4,3\\
3EG J2321-0328&J2323-0317&1.411&28.12&24.13&$<$16.78&0.0&\nodata&0.79&0.87&1.36&4,2,4,3\\
3EG J2358+4604&J2354+4553&1.99&28.45&23.74&17.25&0.34&1.38&0.96&0.74&0&4,1,4,3 \\
3EG J2359+2041&J0003+2129&0.45&26.25&21.58&15.67&-0.6&1.09&0.93&0.71&0.74&1,1,1,3\\
\enddata     
\tablerefs{(1) Sowards-Emmerd {\it et al.} 2003 (2)Sowards-Emmerd {\it
    et al} 2004   2004 (3) Hartman {\it et al.} 1999 (4)Hewitt \& Burbidge 1989
  (5)Veron-Cetty \& Veron 2001  (6)Bloom {\it et al} 2004 (7) NED
  (8) Monet {\it et al.}  (9) Stocke {\it et al.} 1991 (10) Combi {\it et al.} (11)
  Halpern {\it et al.} 2003  (12) Landt {\it et al.} 2001 (13)
  Lauberts \& Valentijn 1989 (14) Wisotzki {\it et al.} 2000 (15)
  Wright \& Otrupcek 1990  (16) Fouque {\it et al.} (17) Liang \& Liu 2003}
\end{deluxetable}
\begin{deluxetable}{lllccccccc}
\tabletypesize{\scriptsize}
\tablecolumns{10}
\tablewidth{0pt}
\tablenum{2}
\tablecaption{Correlation and Regression Analysis for EGRET Blazars}
\tablehead{\colhead {Ind. Var.}& \colhead {Dep. Var.}&\colhead {N}
&\colhead {$\tau$}&\colhead{Prob.}&\colhead{$\rho$}&\colhead{Prob.}&\colhead{ Technique}&\colhead {Slope}&\colhead{Intercept} \\
\colhead{(1)} &\colhead{(2)} &\colhead{(3)} &\colhead{(4)}
&\colhead{(5)} &\colhead{(6)}&\colhead{(7)}& \colhead {(8)} &
\colhead{(9)}& \colhead {(10)}\\}
\startdata
log $L_r$& log $L_{\gamma}$ &122&0.628 & $1 \times 10^{-10}$ &
0.806 & $1 \times 10^{-10}$ & EM & 0.766 &-4.81 \\
log z& log $L_r$ & 122 &0.650 & $1 \times 10^{-10}$ &0.825  &$1
\times 10^{-10}$  & EM &2.47 &27.7 \\
log z& log $L_{\gamma}$&122&0.569 & $1 \times 10^{-10}$ &0.705
&$1 \times 10^{-10}$ &EM &2.12 & 16.4\\
log $L_o$  &log $L_{\gamma}$ & 122&0.299&$1 \times 10^{-6}$ &0.437
& $2 \times 10^{-6}$ & SB & 0.566 & 2.71\\
log z & log $L_o$ & 122 &0.404 & $1 \times 10^{-10}$ &0.564 & $ 7
\times 10^{-10}$& EM&1.26 & 23.7\\
$\alpha_{ro}$&$\alpha_{o \gamma}$&122&-0.556 &$1 \times
       10^{-10}$&-0.701&$1 \times 10^{-10}$ &BJ&-0.519&1.23\\
log$L_o$ & $\alpha_{ro}$ &122 & -0.094& $0.122$&-0.111&0.222&SB&-0.007 &0.892 \\ 
log $L_o$ &$\alpha_{o \gamma}$&122 &0.214& $7 \times
10^{-4}$&0.309& $7. \times 10^{-4}$ &SB&0.032&0.113\\
\enddata     
\end{deluxetable}
\begin{deluxetable}{lllccc}
\tabletypesize{\scriptsize}
\tablecolumns{6}
\tablewidth{0pt}
\tablenum{3}
\tablecaption{Partial Correlation Analysis for EGRET Blazars}
\tablehead{\colhead {Ind. Var.}& \colhead {Dep. Var.}&\colhead {Third
    Var.} &\colhead {N}&\colhead {$\tau$} &\colhead{Prob.} \\
\colhead{(1)} &\colhead{(2)} &\colhead{(3)} &\colhead{(4)}
&\colhead{(5)} &\colhead{(6)}\\}
\startdata
log $L_r$& log $L_{\gamma}$ & log z  &122& 0.341 & $1 \times 10^{-10}$ \\
log $L_o$  &log $L_{\gamma}$ & log z &122&-0.089 &0.148 \\
$\alpha_{ro}$& $\alpha_{o \gamma}$ & log $L_o$ & 122& 0.550&$2 \times
10^{-8}$\\
\enddata     
\end{deluxetable}
\begin{deluxetable}{lllccccccccccc}
\tabletypesize{\scriptsize}
\tablecolumns{14}
\tablewidth{0pt}
\tablenum{4}
\tablecaption{Parameters for Monte Carlo Simulations}
\tablehead{\colhead {Identifier}& \colhead {$log L_{1}$}&\colhead
  {$log L_{2}$}& \colhead {g}&\colhead { $\Gamma_1$} &\colhead{
    $\Gamma_2$}&\colhead{s} &\colhead {$\tau $}&\colhead {$z_{min}$}
&\colhead {$z_{max}$}&\colhead {$\rm h$} &\colhead {$K_{min}$} &\colhead {$K_{max}$}
&\colhead  {$\rm slope$}\\
\colhead{(1)} &\colhead{(2)} &\colhead{(3)} &\colhead{(4)}
&\colhead{(5)} &\colhead{(6)}&\colhead{(7)} & \colhead {(8)} &\colhead
{(9)} &\colhead {(10)} &\colhead {(11)} &\colhead {(12)} &\colhead {(13)} &\colhead {(14)}\\ }
\startdata
Lister SSC& 22.4 & 26.0 & 2.5& 1.001& 25.0&1.5 &0.26&0.0& 4.0& 0.65 &$5 \times 10^{-14}$ & $1.0 \times 10^{-11}$ &1.6\\
Lister ECS & 22.4 & 26.0 &2.5 &1.001& 25.0& 1.5 &0.26&0.0 & 4.0& 0.65& $1.0 \times 10^{-16}$ &$ 1.0 \times 10^{-15}$ & 2.3\\
Bloom SSC & 21.5 & 30.0 & 2.0& 1.001& 30.0 & 1.5 & 0.26& 0.0&5.0 &1.0& $5 \times 10^{-14}$ & $1 \times 10^{-11}$ & 1.6\\
Bloom ECS & 21.3 & 30.0 & 2.0 &1.001& 30.0 & 1.5 & 0.26& 0.0&5.0& 1.0& $4.8 
\times 10^{-17}$ & $5 \times 10^{-17}$& 2.3\\
\enddata     
\end{deluxetable}
\begin{deluxetable}{lllcccccccc}
\tabletypesize{\scriptsize}
\tablecolumns{11}
\tablewidth{0pt}
\tablenum{5}
\tablecaption{Kolmogorov-Smirnov Tests for Monte Carlo Simulations}
\tablehead{\colhead {Identifier}& \colhead {$D_{Lr}$}&\colhead
  {Prob}& \colhead {$D_{L,\gamma}$}&\colhead {Prob} &\colhead{
    $D_z$}&\colhead{Prob} &\colhead {$D_{fr} $}&\colhead {Prob}
&\colhead {$D_{f\gamma}$}&\colhead {Prob}\\
\colhead{(1)} &\colhead{(2)} &\colhead{(3)} &\colhead{(4)}
&\colhead{(5)} &\colhead{(6)}&\colhead{(7)} & \colhead {(8)} &\colhead
{(9)} &\colhead {(10)} &\colhead {(11)}\\}
\startdata
Lister SSC&0.516 &$1 \times 10^{-15}$ &0.172& 0.054 
&0.328 & $4 \times 10^{-6}$&0.303& $3 \times 10^{-5}$& 0.148
& 0.140\\
Lister ECS & 0.426 & $5 \times 10^{-10}$ &0.148 & 0.398& 0.271&0.004 &0.156 &0.103 & 0.189& 0.026\\
Bloom SSC  & 0.205 & 0.012 & 0.230& 0.003&
0.180 & 0.038& 0.205&0.012& 0.164& 0.075 \\
Bloom ECS  & 0.164 & 0.075&0.246 & 0.001& 0.213& 0.008& 0.271& 0.003& 0.197& 0.018\\
\enddata     
\end{deluxetable}
\end{document}